\RequirePackage{fix-cm}
\documentclass[smallextended]{svjour3}       % onecolumn (second format)
\smartqed  % flush right qed marks, e.g. at end of proof
\usepackage{graphicx}
\usepackage{amsmath}
\usepackage{booktabs}
\usepackage{float,lscape}
\usepackage{multirow}
\usepackage{multicol}
\usepackage[]{algorithm2e}
\usepackage{tikz}
\usepackage{lineno,hyperref}
\usepackage{changepage}
\begin{document}

\title{Optimal survival trees ensemble%\thanks{Grants or other notes
}


\author{Naz Gul*\and
	Nosheen Faiz\and
	Dan Brawn\and
	Rafal Kulakowski\and
     Zardad Khan \and
	Berthold Lausen **
}


\institute{N Gul* \at
	 Department of Mathematical Sciences, University of Essex, Colchester CO4 3SQ, UK\\       
	\email{ngul@essex.ac.uk}            \\
	\emph{Department of Statistics, Abdul Wali Khan University, Mardan, Pakistan} \\
		\and
	D. Brown \at
     Department of Mathematical Sciences, University of Essex, Colchester CO4 3SQ, UK \\
	\and
	Z. Khan \at
	Department of Mathematical Sciences, University of Essex, Colchester CO4 3SQ, UK\\       
	\emph{Department of Statistics, Abdul Wali Khan University, Mardan, Pakistan} \\
	\and
	B. Lausen** \at
	\email{blausen@essex.ac.uk}\\
	Department of Mathematical Sciences, University of Essex, Colchester CO4 3SQ, UK \\   
}

\date{Received: date / Accepted: date}

\maketitle

\begin{abstract}
%Following the notion of selecting accurate and diverse trees based on individual and collective performance in an ensemble has recently been studied for classification and regression problems, the possibility of growing a forest of optimal survival trees is considered in this work. 
Recent studies have adopted an approach of selecting accurate and diverse trees based
on individual or collective performance within an ensemble for classification and regression problems. This work follows in the wake of these investigations and considers the possibility of growing a forest of optimal survival trees.
Initially, a large set of survival trees are grown using the method of random survival forest. The grown trees are then ranked from smallest to highest value of their prediction error using out-of-bag observations for each respective survival tree.
% A selected number of the 
The top ranked survival trees are then assessed for their collective performance as an ensemble. This ensemble is initiated with the survival tree which stands first in rank, then further trees are tested one by one by adding them to the ensemble in order of rank. A survival tree is selected for the resultant ensemble if the performance improves after an assessment using independent training data. This ensemble is called an optimal survival trees ensemble (OSTE). The proposed method is assessed using 17 benchmark datasets and the results are compared with those of random survival forest, conditional inference forest, bagging and a non tree based method; the Cox proportional hazard model. In addition to improved predictive performance, the proposed method reduces the number of survival trees in the ensemble as compared to the other tree based methods. 
%Furthermore, the 
 The method is implemented in an $R$ package called ``OSTE''.
\end{abstract}
\section{Introduction}
Survival Analysis concerns the estimation of time until the occurrence of an event of interest.
% is the main objective. 
For example, an event of interest could be the follow up time of an individual from entry into a study until recovery from a certain type of disease.
% could be an event of interest. 
Survival analysis not only estimates and compares survival probabilities of individuals from a specific time point to the endpoint of interest but also finds the ratio of the probability density function to the survival function and assesses the relationship between explanatory variables and survival time.

The presence of unobserved events i.e. censoring reduces the predictive performance of survival modelling. To address this issue, many techniques are developed in the literature. Kaplan and Meier \cite{meier1958nonparametric} computed successive probabilities of non-occurrence of an event at certain points in time, then obtained the product of these probabilities and earlier similarly calculated probabilities for a resultant estimate. Nelson-Aalen \cite{borgan1997three} used the estimation of the cumulative hazard function for censored data. Cox \cite{david1972regression} introduced a large family of semi-parametric models by focusing on hazard functions via the Cox proportional hazard models.  
%The main objective of monitoring and analyzing survival data for long period of time is to estimate the time of occurrence of a particular event of interest in the best way. 
In the search for increased predictive performance in survival models, 
tree based approaches offer one of many attempts. A stand-alone survival tree provides some advantages in respect to interpretation and requires fewer assumptions than simple regression techniques.
 
The main aim in the process of risk prediction model building, is to build a model that predicts future risk accurately. Usually, a model is fitted using given data and then  the performance of the model is checked using same data. However, using the same dataset for constructing a model and for assessing its performance creates over-optimization problems with low generalizability \cite{zhou2012ensemble}. In any model based upon decision trees this level of over-optimisation becomes high due to large-scale searching in each and every node of a tree \cite{breiman1996heuristics}. To alleviate this problem, the given data may be partitioned into training and testing data sub-sets. In this approach, a training part is used for building a model and a testing part is then used for assessing the performance of the derived model \cite{zhou2012ensemble}. This technique decreases sample size and hence the model power is also decreased.
 
 Alternatively, a model is built on a number of random bootstrap samples drawn from the given data. The model performance is then tested using the same data by reporting means and standard deviations.  
 
 Random survival forest is one of these aggregation methods which draws multiple bootstrap samples from the given data. In addition to that, to construct each tree node, a random sample of the independent variables is selected and only these are used in the construction of the tree. Combining the results from these trees i.e. forming an ensemble of survival trees, increases the predictive power of the decision model. Similarly, a survival forest consisting of accurate and diverse survival trees may perform better than a forest grown by combining simple survival trees. Breiman's \cite{breiman1984classification} forest of regression and classification trees also holds to this intuition, which further led to the method of optimal trees ensemble (OTE) for regression and classification \cite{khan2016ensemble1,khan2016ensemble2}.

%Ensemble of optimal trees refines the idea of bagging and random forest based on the Breiman's upper bound for the overall prediction of random forest given as: 
%$$ (PE)^*\leq{\bar{\rho}} {PE_j},$$  where $PE^*$ denotes the overall prediction error of random forest, $\bar{\rho}$ is the weighted correlation between residuals from two independent classification or regression trees and ${PE_j}, j=(1,\dots,B)$ is the prediction error of the $j$th classification or regression tree in the forest. $B$ is the total number of trees in the forest.

OTE combines the best trees i.e. accurate and diverse, from a large number of trees grown initially by Breiman's \cite{breiman1984classification} random forest and thus refines bagging and random forest approaches. On the bases of individual performance with out-of-bag observations, OTE selects a proportion of top ranked trees in the first phase and uses Brier scores to assess these trees on independent training data for their collective performance. For the resultant ensemble, this method selects trees one by one from the trees ordered with respect to the highest prediction accuracy to the lowest prediction accuracy. A tree is discarded, if the predictive performance decreases or does not improve \cite{khan2016ensemble1,khan2016ensemble2}. OTE uses fewer trees and gives comparable results to some other state-of-the-art methods.
% A tree is to be select for the resultant ensemble if its addition to the previously added trees increase its predictive performance \cite{khan2015ensemble}. OTE method select trees for final ensemble one by one starting from the tree with the highest prediction accuracy. The method has been shown to give comparable results using fewer trees than some other state-of-the-art methods considered.

%Here it is aimed to extend 
The extension of OTE to survival data is the main aim for this paper. The objective is to select the best survival trees, in terms of their individual and collective predictive accuracy and integrate them together to develop a new ensemble. This ensemble will be called an optimal survival trees ensemble (OSTE). The results from OSTE are compared with those of Cox proportional hazard model, bagging survival trees, random survival forest and conditional inference forest, using 17 benchmark survival datasets.
\section{Optimal ensemble of survival tree (OSTE)}
In survival analysis an improvement to a single base model , namely a survival tree, is bagging survival trees \cite{hothorn2004bagging}, which combines a number of single survival trees by using the aggregated Kaplan–Meier curve for each new observation.
\textbf{%Predicted survival probability functions of censored event free survival are improved by bagging survival trees. We suggest a new method to aggregate survival trees in order to obtain better predictions for breast cancer and lymphoma patients. A set of survival trees based on B bootstrap samples is computed. We define the aggregated Kaplan–Meier curve of a new observation by the Kaplan–Meier curve of all observations identified by the B leaves containing the new observation. The integrated Brier score is used for the evaluation of predictive models. 
}

 Random survival forest \cite{ishwaran2008random} broadens this idea by selecting a subset of features instead of choosing from the whole set of features while splitting the nodes of the tree.
% Breiman proposed the idea of random forest to further improve bagging by inducing additional randomness in the base model, tree. This is done  Hothorn et al. \cite{hothorn2004bagging} and  these ideas to the analysis of right censored data as bagging survival trees and random survival forest respectively after the use of survival tree structure for survival data \cite{ciampi1981approach}.
 OSTE is an attempt to refine this idea by assessing survival trees both on their collective and individual performance. 

To obtain the ensemble of optimal survival trees divide the given training data $\mathcal{L} = (\mathbf{X},\mathbf{Y})$ randomly into two non overlapping parts $\mathcal{L}_{B} = (\mathbf{X}_{B},\mathbf{Y}_{B})$ and $\mathcal{L}_{V} = (\mathbf{X}_{V},\mathbf{Y}_{V})$. After partitioning, draw $B$ bootstrap samples from $\mathcal{L}_{B} = (\mathbf{X}_{B},\mathbf{Y}_{B})$. Grow a survival tree on each sample by randomly selecting a subset of $p < d$ features at each node of the tree to induce additional randomness. Some of the observations are left out of samples during bootstrapping which are called out-of-bag (OOB) observations. These observations play no role in training the corresponding model, however, they could be used as test data to determine a prediction error for the corresponding individual survival tree. According to their C-index (introduced in Section \ref{cindex}) the grown survival trees are arranged in ascending order and the top $M$ trees are selected. The selected trees are tested one by one for diversity as follows:
\begin{itemize}
	\item The collective performance of survival trees is assessed on independent training data $\mathcal{L}_{V} = (\mathbf{X}_{V},\mathbf{Y}_{V})$ to get the resultant ensemble of survival trees. From the arranged list of trees, an ensemble of two i.e. the second best survival tree combined with the top best survival tree is assessed by using the training data $\mathcal{L}_{V} = (\mathbf{X}_{V},\mathbf{Y}_{V})$. Similarly, in next step, the third best survival tree is added to the ensemble of size two, obtained in the previous step, and the performance is again measured. The decision to add or discard a tree in the final ensemble is dependent on whether or not the prediction error of the ensemble is  decreased or increased respectively. The same steps are repeated for all $M$ survival trees.
	\item A survival tree, $\hat{L}_{k}$ where $k=(1,2,\dots,M)$ is chosen as an optimal survival tree for the intended ensemble of optimal survival trees if its addition to the ensemble with out the $k^{th}$ survival tree fulfils the following criterion
\end{itemize} 
\begin{equation*}
IBS^{(k-)} > IBS^{(k+)},
\end{equation*}
where $ IBS^{(k+)}$ is the integrated brier score (IBS) of the ensemble including the $k^{th}$ tree and $IBS^{(k-)}$ is the integrated brier score of the ensemble in which the $k^{th}$ tree is not yet included. 

\subsection{Concordance index}\label{cindex}
%Discrimination and Calibration \cite{d1997measures} are two measures used to check the predictive performance of mathematical model having binary outcomes. Discrimination check that how much ability the model has to classify the subject into relevant class while calibration describe that how closely the actual outcomes relate to the numerical values of the predicted probabilities. Discrimination is the most preferred one due to the fact that re-discrimination is not possible unlike calibration. On the other hand implementation of discrimination does not affect the calibration. 

In survival analysis, for each subject under study, there is a survival time and prediction of it. Therefore, for the evaluation of the survival model, instead of the absolute survival time for each subject the relative risk of an event for different subjects is considered.
%unlike logistic regression where each subject has to fall into one of two possible categories. Therefore, measuring discrimination in survival analysis is more difficult and so the survival model can be evaluated by considering the relative risk of an event for different subjects instead of the absolute survival times for each subject. 
The concordance index (C-index) \cite{harrell1982evaluating}, an extension of the concept of the receiver operating characteristic (ROC) curve area, is one of the most reported measures used for checking survival model performance. 

The C-index is actually the relative frequency of concordant pairs.
%, \textbf{a pair in which an individual shows high predicted risk for a short survival time.} 
In other words, a pair of two observations is said to be concordant if the observation which is predicted to have an earlier outcome is observed to fail earlier than the other.
%It is where ROC measure the discriminative ability of the under discussion biomarker at each time point \cite{mayr2014boosting}. Concordance index is one of the most reported metrics to predict biomarkers in survival setting.
 
% It is also used for the comparison of models derived in different statistical cultures, such as a Cox regression model and a random survival forest model \cite{bingham2001random}. A pair is said to be concordant if an individual predicted risk is high while the survival time is short. Among all pairs of an individualwhen the data under study is not censored. Briefly, 
The C-index can also be described as the probability that a subject of interest, whose survival time is short is associated with a high value of an indicator (biomarker) and vice versa. In other words C-index measures the discriminative ability of a biomarker. For example, in biomedical research, where grouping of patients into good or poor prognosis groups is required \cite{mayr2014boosting}.

The C-index usually calculated as
\begin{enumerate}
	\item All given observations are paired.
	\item Discard those pairs of observations that have same survival time for an event and those pairs where there is a censored observation at the lower time point. 
	
	\item  Score each of the permissible pairs (the set of remaining pairs) as

	A: 1 if
	\begin{itemize}
		\item A pair consists the observations with unequal survival time and the outcome of an observation with the shorter survival time is correctly predicted.
		\item The survival times are equal, while the event is observed for only one observation with the lower predicted survival time. 
		\item The survival times and their predicted outcomes are equal for both observations.
	\end{itemize}
       B: 0.5 if
       \begin{itemize}
       	\item The outcome of the observations are predicted to be equal for their unequal survival time.
       	\item  The outcome of the observations are predicted to be unequal for their equal survival time.
       	\item For equal survival times the survival time is predicted to be lower for the observation for whom the event remains unobserved. 
       \end{itemize}
   \item The error rate is then calculated using the formula $Error=1-C$ where $C$= Concordance.
\end{enumerate}
%subjects with small survival times and subjects with large survival times.
% under study with a small survival time is associated with a high value of an indicator (biomarker) and vice versa. In other words C-index measures the ability of a biomarker to discriminate between subjects with small survival times and subjects with large survival times. This strategy is very helpful in the field of biomedical research where patients are needed to be subdivided into groups with good or poor prognosis \cite{mayr2014boosting}. It is also applicable to continuous, ordinal and dichotomous outcomes.
  In right censored survival settings, the C-index is defined as:\\ 
\begin{equation}
C = P (\delta_1 >\delta_2 | T_1<T_2),
\end{equation}
where $\delta_1, \delta_2$ are the predicted biomarker values while, $T_1,T_2$ represent event times\cite{mayr2014boosting}. The biomarker does not perform well for $C=0.5$. 
%where $T_1,T_2$ and $\delta_1, \delta_2$ are the event times and the predicted biomarker values \cite{mayr2014boosting}, respectively. The biomarker value shows closeness to a perfect discriminatory power for $C= 1$ while for $C=0.5$ a marker does not perform well.
\subsection{Integrated Brier score (IBS)}
In the context of survival analysis, the squared difference between the survival function indicator and the predicted survival probability is called Brier score. In case of non-censored data, it can easily be calculated by averaging the squared distances between the survival function indicator for the subject and the predicted survival probability given by the model for that subject. \cite{tsouprou2015measures}  In other words, for non-censored data the prediction error is assessed by taking average of the squared residual
  $(\text{observed status} - \text{ predicted status})^2$.
 However, for right censored data, which is the main concern in the OSTE method, the squared residual needs to be weighted at each time point, t. Hence, the integrated brier score (IBS) technique is used. IBS is simply the integration of the Brier score. 
 
%IBS is calculated simply by taking the integration of Brier score. Brier
%score (BS) is actually a measure of the mean squared difference between the actual outcome $y_{i}$ and the predicted probability of the possible outcome for the $i$th observation. The advantage of BS over other common methods for prediction assessment is its correct classification in different outcome groups and agreement of the predictions with the true risk. In other words, these are called discrimination and calibration \cite{tsouprou2015measures}.
%
%In the context of survival analysis the Brier score is the squared difference between the survival function indicator and the predicted survival probability. In free censored survival data BS can easily be estimated by taking average of the squared distances between the indicator of survival function for the subjects and the predicted survival probability given by the model for that subjects \cite{tsouprou2015measures}. However, for right censored data which is the main concern here in this thesis, the integrated brier score technique is used. 

Let $T_{i}$ be the time of the event of interest of subject $i$ and $\delta_{i}= I(T_{i} > t_{0})$ be the outcome while the estimated probability of a subject at risk surviving beyond $t_{0}$( the follow-up time) is $\hat{S}(t_{0}|\mathbf{x}_{i})$
%  and $\mathbf{x}_{i}$ are the given covariates 
  then the BS is given as 
  %\cite{tsouprou2015measures}
\begin{eqnarray*}
	BS (t_{0}) &=& E(\textbf{I}(T_{i} > t_{0}) - \hat{S}(t_{0}|\mathbf{x}_{i}))^2,\\
	&=& E(\textbf{I}(T_{i} > t_{0}) - S(t_{0}|\mathbf{x}) -(\hat{S}(t_{0}|\mathbf{x}_{i})-S(t_{0}|\mathbf{x})))^2,\\
	&=&  E(\textbf{I}(T_{i} > t_{0}) - S(t_{0}|\mathbf{x}))^2 + E(\hat{S}(t_{0}|\mathbf{x}_{i})- S(t_{0}|\mathbf{x}))^2.
\end{eqnarray*}
%The censored indicator $\delta_{i}$ is not always easy to calculate. If an individual $i$ survived at least until time $t_{0}$ i.e $\textbf{I}(T_{i} > t_{0})$ then  $\delta_{i} = 1$ otherwise $0$.
For an individual that is censored before $t_{0}$ i.e $T_{i} < t_{0}$, the observation is consider to be unknown.
% The indicator is consider to be unknown if the individual is censored before $t_{0}$ i.e $T_{i} < t_{0}$ but $o_{i}=0$. 
 To overcome this issue, Inverse-probability-of-censoring weighting (IPCW) \cite{robins2000correcting} is used .
Hence, the Brier score can be written as:
\begin{equation*}
\hat{BS}(t_{0}) = \frac{1}{n} \sum\limits_{i=1}^{n} (\textbf{I}(T_{i} > t_{0}) - \hat{S}(t_{0}|x_{i}))^2 w_{i},
\end{equation*} 
where 
\begin{equation}
w_{i}=\left\{
\begin{array}{@{}ll@{}}
0, &\text{  if  }  \;T_{i} > t_{0} \;\text{ and }  \;\delta_{i} = 0, \\ 
\frac {1}{\hat{G}(t_{0})}, & \text{  if  } \;T_{i} > t_{0},\\ 
\frac {1}{\hat{G}(T_{i})}, & \text{  if  }  \; T_{i} < t_{0} \;\text{ and } \;\delta_{i} = 1,
\end{array}\right.
\end{equation}
where $\hat{G}(t_{0})$ is the Kaplan-Meier estimate of the probability of being uncensored at time $t_{0}$. 
%The predictive performance of the model is said to be 
%better for the lower values of the Brier score. 0 indicates perfect predictions, however, in practice it is very rare or may be impossible to get 0 value for the brier score.The Brier score defined above is a function of time $t_{0}$ where 
In right censored data over a range of time points a measure of predictive accuracy is required. Therefore, an integrated Brier score (IBS) with respect to some weight function for ${t \in (0,t^*)}$ can be estimated by integrating the BS calculated via the method shown above %\cite{rahman2012validation}

%The integrated Brier score (IBS) for ${t \in (0,t^*)}$ is calculated as follows:

\begin{equation*}
IBS(t_{0}) = \int_{0}^{t^*}\hat{BS}(t_{0}) dt \; \hat{w}_{(t_{0})},
\end{equation*} 
where a weight function at individual time points is represented by $ \hat{w}_{(t_{0})}$, integration of the area under the prediction curve.
% by using trapezoidal rule (The rule of approximating the region under the graph of the function as a trapezoid and calculating its area).
.
% \cite{rahman2012validation}.
% The trapezoidal rule works by approximating the region under the graph of the function as a trapezoid and calculating its area.that integrate the area under the prediction curve \cite{rahman2012validation}. 
 
\subsection{OSTE Algorithm}
The algorithm of the proposed method OSTE consists of the following steps:
\begin{itemize}
	\item Partition the training data randomly into two non-overlapping parts  $\mathcal{L}_{B}$ and $\mathcal{L}_{V}$ i.e $\mathcal{L} = (\mathbf{X},\mathbf{Y})$=$\mathcal{L}_{B}$ and $\mathcal{L}_{V}$
	\item Draw $B$ bootstrap samples from the data $\mathcal{L}_{B} = (\mathbf{X}_{B},\mathbf{Y}_{B})$.
	\item On the bootstrap samples grow survival trees in such a way that at each node $p < d $ features are chosen.
	\item On the bases of individual prediction error using OOB data, arrange the grown trees in ascending order and chose the top $M$ trees. The prediction error is estimated via the concordance index given in Section \ref{cindex}.
	\item Add the $M$ selected trees one by one starting from the single top tree and calculate the integrated Brier score. Select the survival tree if the result is improved after testing performance on the validation data  $\mathcal{L}_{V} = (\mathbf{X}_{V},\mathbf{Y}_{V})$.

%	Check the performance on validation data  $\mathcal{L}_{V} = (\mathbf{X}_{V},\mathbf{Y}_{V})$. Select the survival tree if the results are improved otherwise discard.
	\item Predict new data by combining the results of the selected trees into the resultant ensemble.
%	 New data are predicted by combining the results of the selected trees in the final ensemble.  
\end{itemize} 
\subsection{Related work}
In the literature there are many other techniques used by researchers to resolve right censored survival data, for example regularized Cox regression models \cite{wright2017unbiased}. However, these models violate the proportional hazard assumptions and hence, miss-specify the model \cite{wright2017unbiased}. Alternatively, machine learning methods such as Conditional inference forests (CIF), have been suggested as have random survival forest (RSF) and the bagging of survival trees. CIF uses different statistical approaches while selecting the split variables and the split points. This approach creates problems in the case of non-linear covariates. Maximally selected rank statistic \cite{lausen1992maximally} is used to avoid this difficulty and to reduce bias \cite{wright2017unbiased}.
The proposed statistic is applied to obtain an exact p-value.
% for classification. It also asses the effect of selected cut-points in case of binary class problems \cite{lausen1992maximally}. 
%Generalized maximally selected rank statistic is used to evaluate large number of cut points and to analyze the asymptotic distribution of these maximally selected statistics \cite{hothorn2008generalized}. 
Moreover, to get the exact distribution of maximally selected rank statistics a lower bound is calculated by extending an algorithm originating from linear rank statistics \cite{hothorn2003exact}.

	In the literature, three types of artificial neural network (ANN) are proposed for survival analysis problems. These are the prediction of the survival time of a subject from the given data directly, neural network survival analysis has been employed \cite{dunn2009basic}, where the survival status of a subject has been taken as the output of the neural network \cite{dunn2009basic} and the extension of the Cox PH model  \cite{faraggi1995neural} to the non-linear ANN predictor with a suggestion for fitting of a neural network.
%	 in models are considered to be the third type.  

%In the context of clinical prediction two approaches Naive Bayes (NB) and Bayesian network (BN) \cite{friedman1997bayesian} that result the probability of the event of interests are commonly used. A Naive Bayes, an effective prediction algorithm, is used for the prediction in clinical medicine \cite{bellazzi2008predictive}. Similarly, Bayesian network, showed the  graphical representation of a theoretical distribution over a set of variables. The visual representation of all the relationships between the variables make it interpretable. Recently, the representation power of this method is combined with the Accelerated Failure Time (AFT) Model \cite{kalbfleisch2011statistical} by estimating the previous probabilities to future time point \cite{ameri2016survival}.

\section{Experiment and results}
\subsection{Benchmark datasets}
%The datasets used for the purpose of benchmarking are called benchmark datasets. In this work, 
To assess and compare the predictive performance of the proposed OSTE approach, with other state-of-the-art methods a total of 17 benchmark datasets are considered. 
%to assess the performance of the proposed method OSTE, in comparison to other state-of-the-art methods. All these datasets are open problems from various sources used to evaluate and compare different learning algorithms. These datasets are summarized briefly
 A brief summary of these chosen datasets are given in Table 1. which details the number of observations and the number of the features the type of the features, whether real, integer or nominal is given against each dataset.
\begin{table}[htbp]
	\centering
	\caption{Datasets description: Number of observations, number of features, type of features whether integer, real, or nominal (I/N/F) and data source are given against each dataset.}
	\begin{tabular}{lccccc}
		\toprule
		Datasets & \multicolumn{1}{c}{No. of} & \multicolumn{1}{c}{No. of} &\multicolumn{1}{c} {Features}& \multicolumn{1}{c}{Censored} & \multicolumn{1}{c}{Source} \\
		& \multicolumn{1}{c}{observations} & \multicolumn{1}{c}{features} &\multicolumn{1}{c} {type}& \multicolumn{1}{c}{observations} & \multicolumn{1}{c}{} \\
		\midrule
		kidney & 119   & 3     & (2/1/0) & 26& \cite{KMsurv} \\
		twins & 24    & 4     & (4/0/0) & 8& \cite{KMsurv} \\
		kidtran & 863   & 5     & (5/0/0) & 140&\cite{KMsurv} \\
		channing & 462   & 5     & (5/0/0) & 176&\cite{KMsurv} \\
		hodg  & 43    & 6     & (6/0/0) & 26&\cite{KMsurv} \\
		myeloid & 646   & 6     & (5/0/1) &320& \cite{survival-package} \\
		veteran & 137   & 8     & (0/7/1) &128& \cite{survival-package} \\
		retinopathy & 394   & 9     & (5/1/3)&155 & \cite{survival-package} \\
		bfeed & 927   & 10    & (10/0/0) &892& \cite{KMsurv} \\
		GBSG2 & 686   & 10    & (7/0/3) &299& \cite{pec} \\
		&	&&&&\url{https://www.ncbi.}\\
		NKI   & 295   & 14    & (0/8/6) &79& \url{nlm.nih.gov/gap/?} \\
		&&&&&\url{term=phs000547.v1.p1} \\
		cgd   & 203   & 15    & (6/4/5) &76& \cite{survival-package} \\
		colon & 1858  & 15    & (0/15/1) &920& \cite{survival-package} \\
		cost  & 518   & 15    & (4/1/10) &404& \cite{pec} \\
		burn  & 154   & 17    & (17/0/0) &99& \cite{KMsurv} \\
		Pbc   & 418   & 19    & (11/7/1) &347& \cite{survival-package} \\
		BMT   & 137   & 22    & (22/0/0) & 81&\cite{KMsurv} \\
		\bottomrule
	\end{tabular}%
	\label{datas}%
\end{table}%

A brief description of these datasets are now follows.

The dataset \textbf{Veteran} has been taken from a randomized trial of two treatment procedures for lung cancer. A total of 137 patients are observed, measuring their survival time since the start of the treatment with a status 1 for dead and 0 for others. Covariates also considered are the type of treatment whether standard or test drug, type of cell, the time since the diagnosis, age, the Karnofsky score. any prior therapy i.e 0 if none and 1 for yes. The survial time has been recorded for days while the time since diagnosis has been recorded for months.

\textbf{Kidtran} dataset is taken from the study designed to assess the time to first clinically apparent infection in a group of patients with renal insufficiency. Total 863 cases are observed on 5 features i.e. gender (male, female), race( white, black), age in years, time which shows period of study, death indicator delta (0 if alive otherwise 1). The original source of the dataset is \cite{klein2005survival}.

\textbf{myeloid} dataset consists of 646 observations with features treatment arm A or B, time to death represented as futime is 1 for a death and 0 for censoring, time to transplant of hematropetic stem cell, time until complete response and time to relapse of disease.

The \textbf{hodg} dataset has 43 observations made on 6 features i.e, graft type 1 for allogenic and 2 for autologous, disease type 1 and 2 for Non Hodgkin lymphoma and Hodgkins disease respectively, time to death, delta (death/relapse indicator), Karnofsky score and waiting time in months to transplant.

The \textbf{retinopathy} dataset is based on a trial to delay diabetic retinopathy through laser coagulation treatment. A data frame consists of 394 observations on 9 variables, type of laser used, treated eye, person age at diagnosis time, type of diabetes, trt that is 0 for control eye and 1 for treated eye, time to loss of vision and eye risk score. A variable status is recorded as censoring indicator.
For each patient there are two observations in the dataset, one for the eye
received laser treatment and the other for the untreated eye. The time when s treatment starts to the time when visual acuity dropped below 5/200 is considered as the event of interest for each eye. The difference between actual time when vision is lost and minimum possible time to event is considered as a survival time.

\textbf{bfeed} dataset consists of breast feeding related information collected from 927 mothers (with first born children) who choose breast feeding. This dataset is actually the main section of the survey conducted by “ the National Longi
tudinal Survey of Youth“ in 1983. The main aim of the survey is to collect information from females about any pregnancies that have occurred in 1983. The response feature in the dataset is duration followed by an indicator feature showing that weaning to infant is completed or not. Other covariates observed during this study are year of child’s birth, race, education, age, poverty, smoking status, alcohol-drinking and lack of prenatal care status in first trimester of pregnancy.

The datasets\textbf{ kidney} and \textbf{cgd} are records of assessment time to first exit site infection in patients with renal insufficiency and time to serious infections observed in granulotomous disease (CGD) through to the end of the study respectively. A total of 119 patients are observed in kidney dataset in two groups whether the catheter is placed surgically or percutaneously with a delta as the censored indicator. This dataset is available free in \texttt{KMsurv} R package  \cite{KMsurv}. On the other side, a total of 203 cases on 15 features are observed in the cgd dataset. These features are enrolling centre, treatment whether placebo or gamma interferon, sex, age, at study entry, height (in cm), weight (in kg), inheritance pattern, use of steroids and prophylactic antibiotics, a categorization of the centres into 4 groups, days to last follow-up, start and end of each time interval and observation number within subject with the status 1 if the interval ends with an infection as censoring indicator. The original source of the data is ``Counting Processes and Survival Analysis``\cite{fleming2011counting}.

Dataset \textbf{twins} is the record of 24 cases who died from coronary heart disease (CHD). The survival time of each individual is recorded in months with the indicator death sets as 1 if the cause of death is CHD.  The other two features show the identification number and gender of an individual under the study.

\textbf{burn} dataset consists of information about the methods used to take care of the burned patients, the infections in their wound and other medical concerns. This dataset consists of the medical records of 154 burned patients treated during the 18-months study period. During study the time until staphylococcus infection was recorded in days with an indicator whether an infection had occurred or not. Gender, race, severity of the burn, burn site and type of burn, time to excision and time to prophylactic antibiotic treatment administration along with the two indicator features, namely, whether the patient’s wound had been removed or not and whether the patient had been treated with an antibiotics during the course of the study.

The information about those women who suffered from breast cancer is collected in \textbf{GBSG2} dataset. Originally the dataset is available in ``Building multivariable prognostic and diagnostic models transformation of the predictors by using fractional polynomials``\cite{sauerbrei1999building}. A total of 686 women are observed on features, age, time of recurrence free survival time (in days), hormonal therapy as a two level factor whether no or yes, menopausal status as a factor at two levels, premenopausal and postmenopausal recorded as horTh and menostat respectively, tumor size and grade, number of positive nodes and progesterone receptor, estrogen receptor with censoring indicator  i.e 0 if censored. The dataset is freely available in the \texttt{pec} R package \cite{pec}.

 The dataset Channing has observations of a total of 462 individuals allowed easy access to medical care without any additional financial burden, as they were covered by a health care program provided by the centre. The original source of the data is the "Survival Analysis Techniques for Censored and Truncated Data`` \cite{klein2005survival}. There are a total of 5 features in the dataset, death status (1 or 0), age of entry into retirement home, age of death or left retirement home, difference between the above two ages and gender.

%%\subsection{rhDNSE}
%%This dataset is the results of a randomized trial of rhDNase for the treatment of cystic fibrosis. In 1992 a randomized double-blind trial was conducted for comparing rhDNase (human enzyme normally present in the mucus of human lungs that digests extracellular DNA) to placebo. Patients were monitored for pulmonary exacerbations, along with measures of lung volume and flow. The primary endpoint was the time until first pulmonary exacerbation; however, data on all exacerbations were collected for 169 days. Total of 767 patients were observed on following 7 variables i.e. enrolling institution, trt (treatment arm: 0=placebo, 1= rhDNase), date of entry into the study, date of last follow-up, forced expriatory volume at enrollment, a measure of lung capacity, days from enrollment to the start of IV antibiotics and days from enrollment to the cessation of IV antibiotics. This dataset is readily available in R package \texttt{KMsurve}  \cite{KMsurv}.

 \textbf{Pbc} dataset consists 424 patients who were eligible for the controlled randomized placebo trial of the D-penicillamine drug. The observed features in this study are alkaline phosphotase, age, serum albumin and serum bilirunbin, presence of ascites, aspartate aminotransferase, serum cholesterol, urine copper, edema  as 0, 0.5 and for no edema, untreated or successfully treated and edema despite diuretic therapy respectively and presence of hepatomegaly or enlarged liver.

\textbf{colon} is the collection of information about colon cancer disease  observed in the first successful adjuvant chemotherapy. Two records per individual i.e, one for recurrence and one for death are recorded. There are a total of 1858 subjects observed on 15 features. These features are study, rx as Treatment - Obs(ervation), Lev(amisole) and Lev(amisole)+5-FU, sex and age of patient, obstruction of colon by tumour, perforation of colon, adherence to nearby organs, number of cancer detectable lymph nodes, days until event or censoring, differentiation of tumour, Extent of local spread, time from surgery to registration, more than 4 positive lymph nodes, event type and censoring status.
%\subsection{BMT}
The \textbf{BMT} dataset consists of information about the recovery process from a bone marrow transplantation of 137 patients. At the time of transplantation several risk factors were measured. For each disease, patients were grouped into risk categories based on their status. 
  Risk factors denoted by z1 to z10 consists of recipient and donor age, gender, cytomegalovirus immune status (CMV) status, waiting time from diagnosis to transplantation, their French-American-British (FAB) classification based on standard morphological criteria, Hospital and MTX as a Graft-Versus-Host- Prophylactic that is 1 if Yes. t1 and t2 represent  time to death and a disease free survival time. d1, d2 and d3 are recorded as death indicator, relapse indicator and disease free survival indicator while da, dc and dp features shows acute and chronic GVHD indicator and platelet recovery indicator respectively. ta, tc and tp show time to acute Graft-Versus-Host and chronic Graft-Versus-Host disease respectively.

In the\textbf{ NKI} dataset the gene expression measurements of 337 lymph node positive breast cancer patients are recorded. The computational burden is reduced and missing data is eliminated by excluding all SNPs with a call fraction below 100\%  keeping 151,346 SNPs. The endpoint relapse-free survival are analysed. This dataset is available at dbGaP and has ID \texttt{phs000547.v1.p1} (\url{https://www.ncbi.nlm.nih.gov/gap/?term=phs000547.v1.p1}).
%\subsection{cost}

The \textbf{cost} dataset contains a subset of the data from the Copenhagen stroke study which observed 518 stroke patients. There are a total of 14 features i.e. age and sex, Hypertension, History of ischemic heart disease at admission, history of previous strokes before admission, history of other disabling diseases, daily alcohol consumption, diabetes mellitus status indicating if the glucose level was higher than 11 mmol/L, daily smoking status, atrial fibrillation, stroke subtype, stroke score, cholesterol level  with the survival time
%(in days) and status (0: censored, 1: event). The dataset is freely available in $R$ package \texttt{pec} \cite{pec} while the original source is ``Evaluating random forests for survival analysis using prediction error curves" \cite{mogensen2012evaluating}.

\subsection{Experimental setup} \label{setup}
%Each data set is divided into training and testing parts.
 A random sub-set of 70\% of each dataset is taken for the training while the remaining 30\% is considered as testing data for all the methods considered in the analysis.
% , the same training and testing parts are used for valid comparison.
 
 In the case of OSTE, an initial ensemble of a total of 1000 independent survival trees is grown on bootstrap samples. At total of 95\% of the training data is used for bootstrapping. At each node of a tree $p$ features are randomly selected from the total of $d$ features as a splitting criterion using the log-rank statistic. The diversity is checked on remaining 5\% of the training data. Usually $p$ i.e $p= \sqrt{d}$, the default value in the standard random survival forest is considered for all datasets. Based on individual accuracy, $20\%$ of total grown trees are selected while keeping the terminal node size fixed at 3. For each dataset a total of 1000 runs is performed and using the remaining 30\% test data, the final results are averaged.
%  from all these runs.

%%In the case of OSTE,  a total of 1000 independent survival trees are grown on bootstrap samples selected from 95\% of the training data as the initial ensemble. For splitting the nodes of the trees, $p$ features are randomly selected at each node from the total of $d$ features while growing the trees. The remaining 5\% of the training data is used for diversity check. For all datasets the number $p$  of feature is taken as the square root of the total features i.e $p= \sqrt{d}$, which is also the default value in the standard random survival forest. Terminal node size is fixed at 3. Based on individual accuracy, $M=20\%$ of total trees grown are selected. A total of 1000 runs are performed for each data set and the final results are the average from all these runs using the 30\% test data. Log-rank statistic is used while growing the forest. 
 For RSF, the number of unproned trees from the set $\{500,1000,1500,2000\}$ is tuned using 10-fold cross validation. The corresponding training part is fine tuned using 10-fold cross validation for all possible values of the number of features variable \texttt{mtry}.   
%In the case of RSF, number of trees is tuned by using 10-fold cross validation on values from the set $\{500,1000,1500,2000\}$. 
%Tree are grown unproned with terminal node size equal to 3. 
.%The value of \texttt{mtry} is tuned by using 10-fold cross validation considering all possible values of the number of features for all the datasets on the corresponding training part
 The Log-rank statistic is used while growing the trees in package \texttt{ranger} \cite{ranger} by keeping terminal node size equal to 3. 
  Unpruned survival trees are grown on $\{500,1000,1500,2000\}$ for the bagging survival trees method using the $R$ package \texttt{ipred} \cite{ipred}. 10-fold cross validation is used for tuning while a log-rank statistic
  is used for measuring the  between nodes distance.
%   a log-rank statistic is used.  
%  The number of trees denoted by \texttt{nbagg} in the $R$ package is the only hyper parameter that is tuned by using 10-fold cross validation on $\{500,1000,1500,2000\}$. Trees are grown unproned. Log-rank statistic is used while growing the forest. 

For conditional inference forest, the r package \texttt{party} \cite{hothorn2006survival} is used. The total number of trees and number of features selected at each node for splitting is denoted by \texttt{ntree} and \texttt{mtry} respectively, these are the only hyper-parameters that are tuned using values from the set $\{500,1000,1500,2000\}$. 

For Cox-proportional hazard model, a non-tree based method, the package \texttt{survival} \cite{survival-package} is used. For all parameters the default values are considered. For \texttt{init} the default is zero for all features. Until the relative change in the log partial likelihood is smaller than $1\mathrm{e}{-09}$, i.e. \texttt{eps}$< 1\mathrm{e}{-09}$ the iterations will remain continue. Value 20 is a default for maximum iteration attempts for convergence. The Efron approximation\cite{efron1977efficiency}  is used to handle ties. 

%For Cox-proportional hazard model, the package \texttt{survival} \cite{therneau2015survival} is used with the default values for all parameters. For the variable initial iteration values, the default is zero for all features, i.e \texttt{init} is set to zero. Iterations continues until the relative change in the log partial likelihood is smaller than $1\mathrm{e}{-09}$, i.e. \texttt{eps}$< 1\mathrm{e}{-09}$. The maximum iteration attempts for convergence is 20 by default. For tie handling, the Efron \cite{efron1977efficiency} approximation is used as the default i.e. \texttt{ties=``efron''}.  
\section{Results and discussion}

Using the experimental settings given in the previous section, the integrated Brier scores for all the methods are calculated on the introduced datasets. The results are given in Table 2. The final results are the average integrated Brier scores from a total of 1000 runs for each of the methods. Each time, given data is randomly divided into training and testing parts as described in the above section. It is clear from the table that OSTE is giving better average results than the alternative methods in 5 out of 17 data sets. On 5 datasets bagging outperformed others while, on 1 data set RSF gives better results. On the twins dataset the results of bagging and RSF are same. CIF gave better results on 4 and Coxph on 3 datasets.  

\begin{table}[htbp]
	\centering
	\caption{Integrated Brier scores of the methods against each data set. The best score is highlighted in bold font.}
	\begin{tabular}{lcccccccc}
		\toprule
		Datasets & \multicolumn{1}{c}{n} & \multicolumn{1}{c}{d}&\multicolumn{1}{c}{E} & \multicolumn{1}{c}{Cox} & \multicolumn{1}{c}{bagging} & \multicolumn{1}{c}{RSF} & \multicolumn{1}{c}{CIF} & \multicolumn{1}{c}{OSTE} \\
		\midrule
		kidney & 119   & 3 &26  & 0.1272 & \textit{0.1296} & \textit{0.1296} & \textbf{0.1257} & 0.1291 \\
		twins & 24    & 4&8     & 0.0144 & 0.0132 & \textbf{0.0131} & 0.0147 & 0.0139 \\
		kidtran & 863   & 5&140     & 0.0341 & 0.0324 &0.0345  &\textbf{ 0.0203} & 0.0357 \\
		channing & 462   & 5&176     & 0.0584 & \textbf{0.0512} & 0.0554 & 0.0664 & 0.0550 \\
		Hodg  & 43    & 6&26     & \textbf{0.1521} & 0.1885 & 0.1836 & 0.1703 & 0.2067 \\
		myeloid & 646   & 6 &320    & 0.1393 & \textbf{0.1348} & 0.1349 & 0.1360 & 0.2474 \\
		veteran & 137   & 8 &128    & 0.2571 & 0.1707 & 0.1692 & \textbf{0.1582} & 0.1683 \\
		retinopathy & 394   & 9&155     & 0.1757 & 0.1795 & 0.1765 & \textbf{0.1714} & 0.1762 \\
		bfeed & 927   & 10&892    & 0.1925 & 0.2397 & 0.1942 & 0.1941 & \textbf{0.1478} \\
		GBSG2 & 686   & 10 &299   & \textbf{0.0148} & 0.0151 & 0.0149 & 0.0182 & 0.0170 \\
		NKI   & 295   & 14 &79   & 0.1510 & 0.1154 & 0.1113 & \textbf{0.1077} & 0.1110 \\
		cgd   & 203   & 15 &76   & 0.2831 & \textbf{0.0819} & 0.0862 & 0.0831 & 0.0844 \\
		colon & 1858  & 15&920    & 0.1737 & \textbf{0.1534} & 0.1605 & 0.1735 & 0.1897 \\
		cost  & 518   & 15&404    & \textbf{0.1764} & 0.1825 & 0.1807 & 0.1851 & 0.1789 \\
		burn  & 154   & 17& 99   & 0.1661 & 0.1477 & 0.1474 & 0.1527 & \textbf{0.1469} \\
		Pbc   & 418   & 19&347    & 0.0669 & 0.0669 & 0.0504 & 0.0523 & \textbf{0.0082} \\
		BMT   & 137   & 22 &81   & 0.0799 & 0.0450 & 0.0560 & 0.0511 & \textbf{0.0299} \\
		\bottomrule
	\end{tabular}%
	\label{ibs}%
\end{table}%

Figures 2-4 give the results in the form of box plots for 17 datasets. Colors brown, gray, red, yellow and blue are used for the box plots of Cox, bagging, RSF, CIF and OSTE receptively. Figure 2 gives the integrated Brier scores for all the methods for the datasets veteran, kidtran, bfeed, twins, GBSG2 and burn. For two datasets, kidtran and bfeed OSTE shows better performance while for other datasets the results of OSTE are similar to the alternative methods.
%\begin{figure}[H] 
%	\centering
%		\includegraphics[width=14cm,height=10cm]{./chapter4/Veteran}
%	\caption{The boxplot showing a comparison of IBS on veteran dataset for all the methods. Cox, bagging, RSF, CIF and OSTE are shown by brown, gray, red, yellow and blue colours, respectively. CIF gives better results than the others. RSF and OSTE give comparable results.}
%	\label{boxveteran}
%\end{figure}

\begin{figure}[H]
	%\begin{figure}[h!]
	\centering
	$\begin{array}{cc}
	\text{ \ \ \ \ }(veteran)&\text{ \ \ \ \ }(kidtran) \\ 
	\includegraphics[width=6cm,height=5cm]{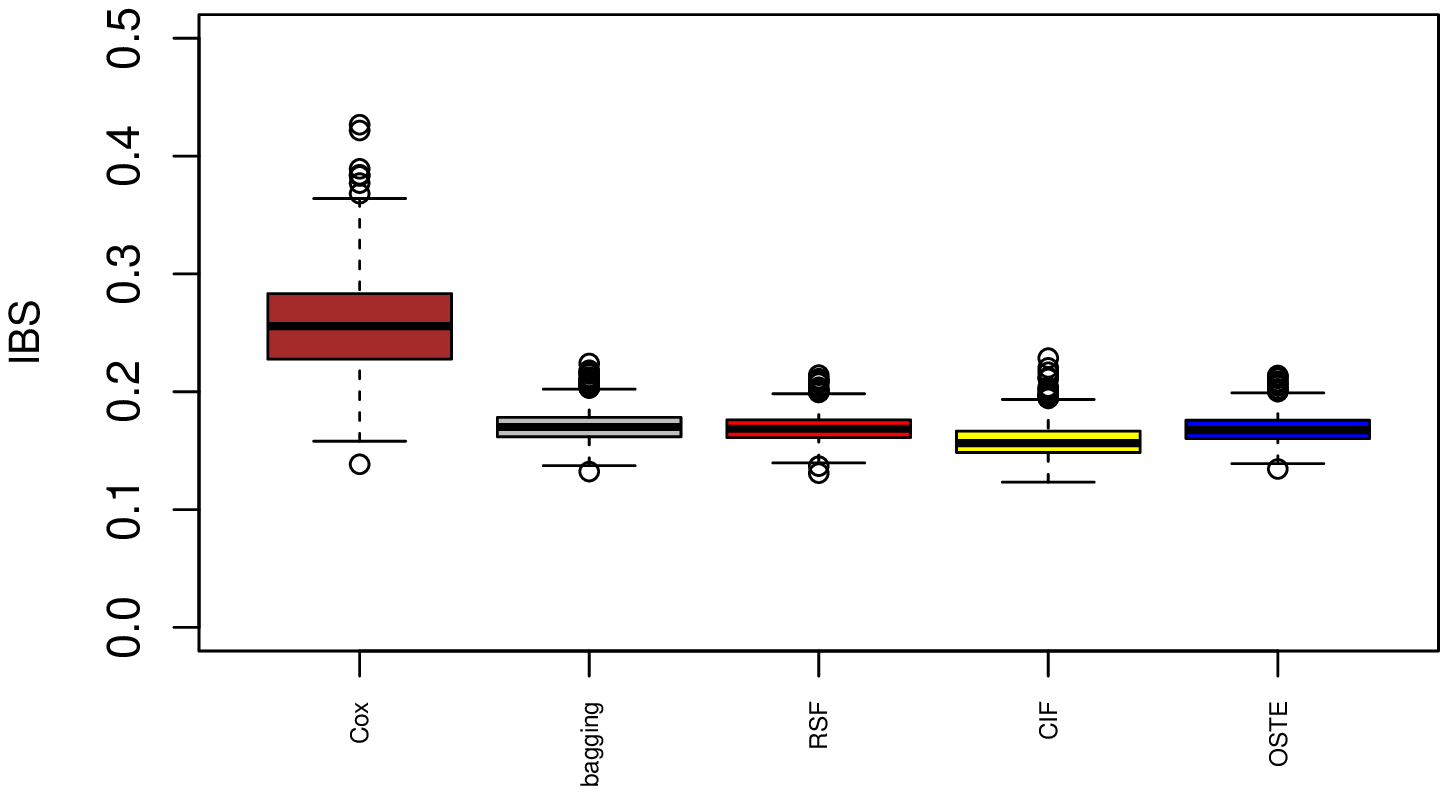} & \includegraphics[width=6cm,height=5cm]{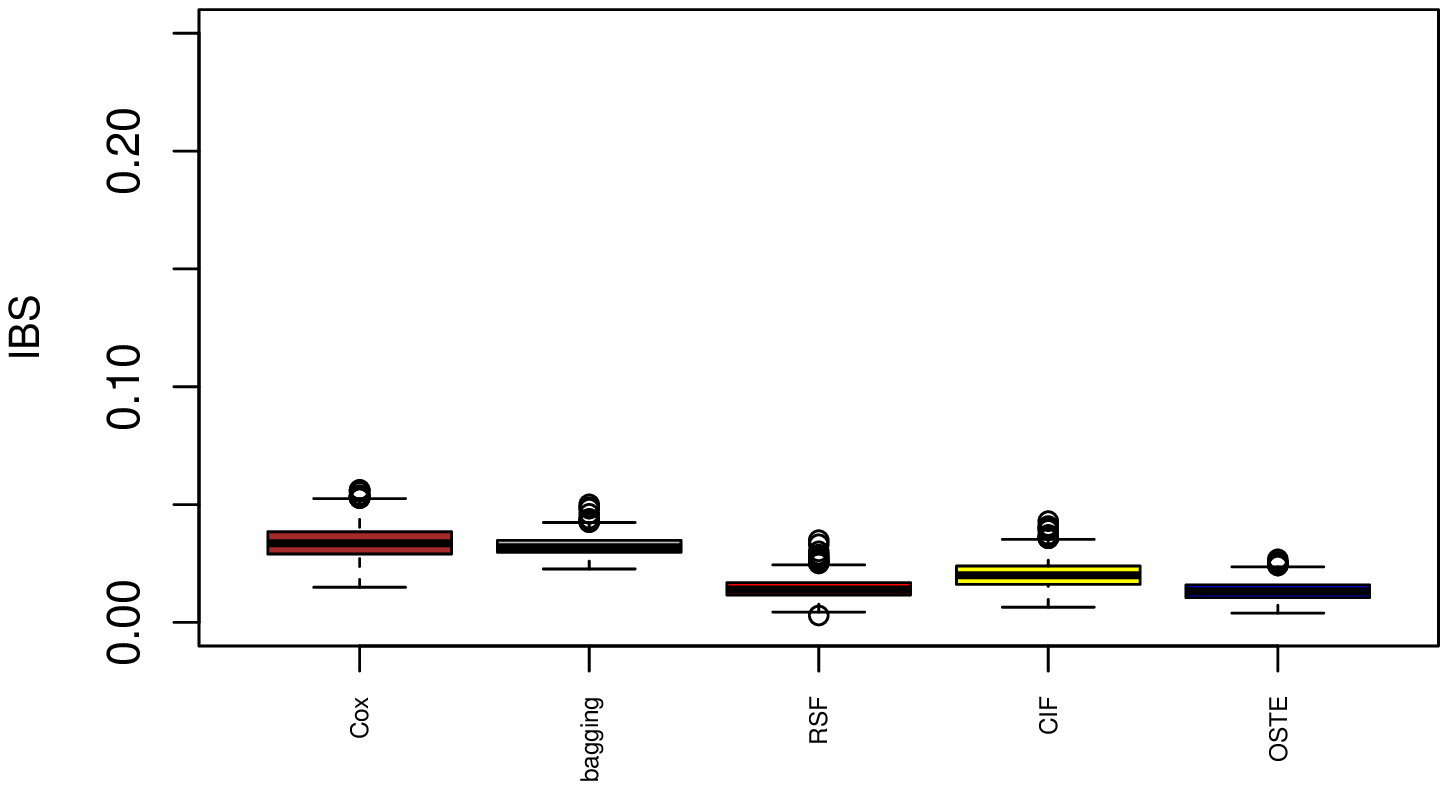}\\
	\text{ \ \ \ \ }(bfeed)&\text{ \ \ \ \ }(twins)\\
	\includegraphics[width=6cm,height=5cm]{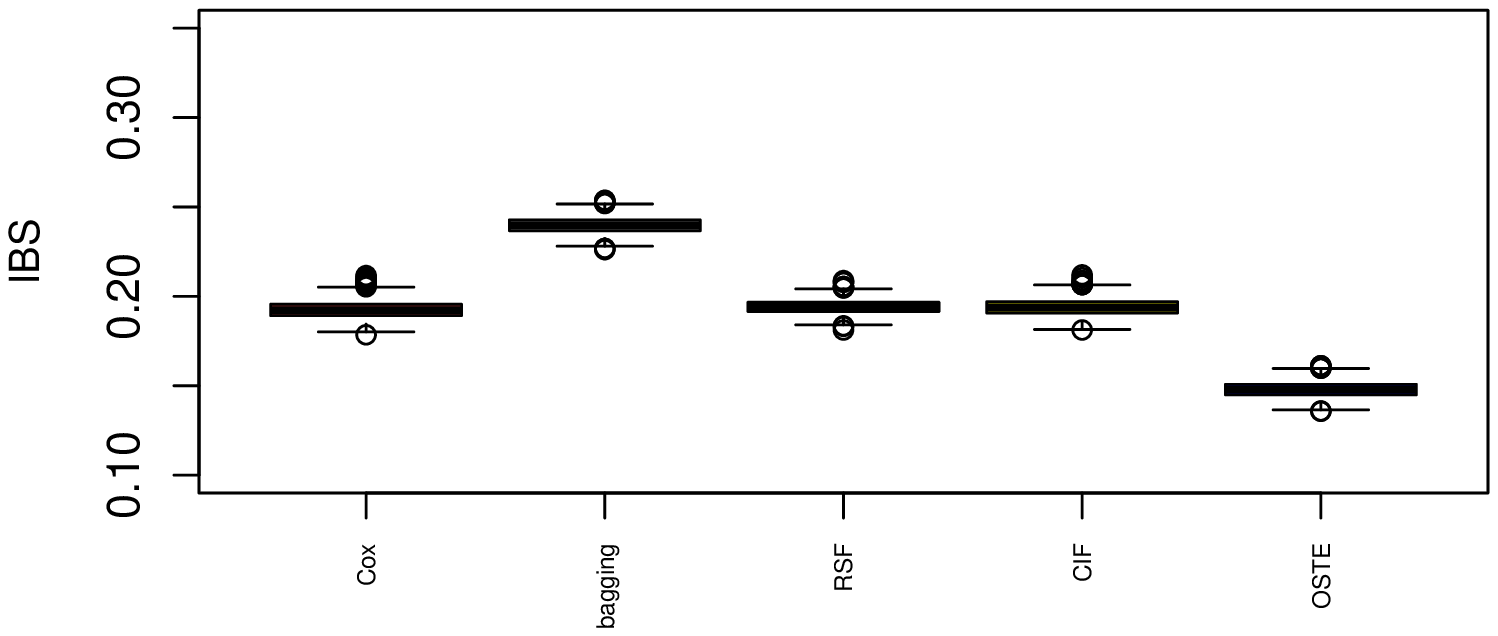} & \includegraphics[width=6cm,height=5cm]{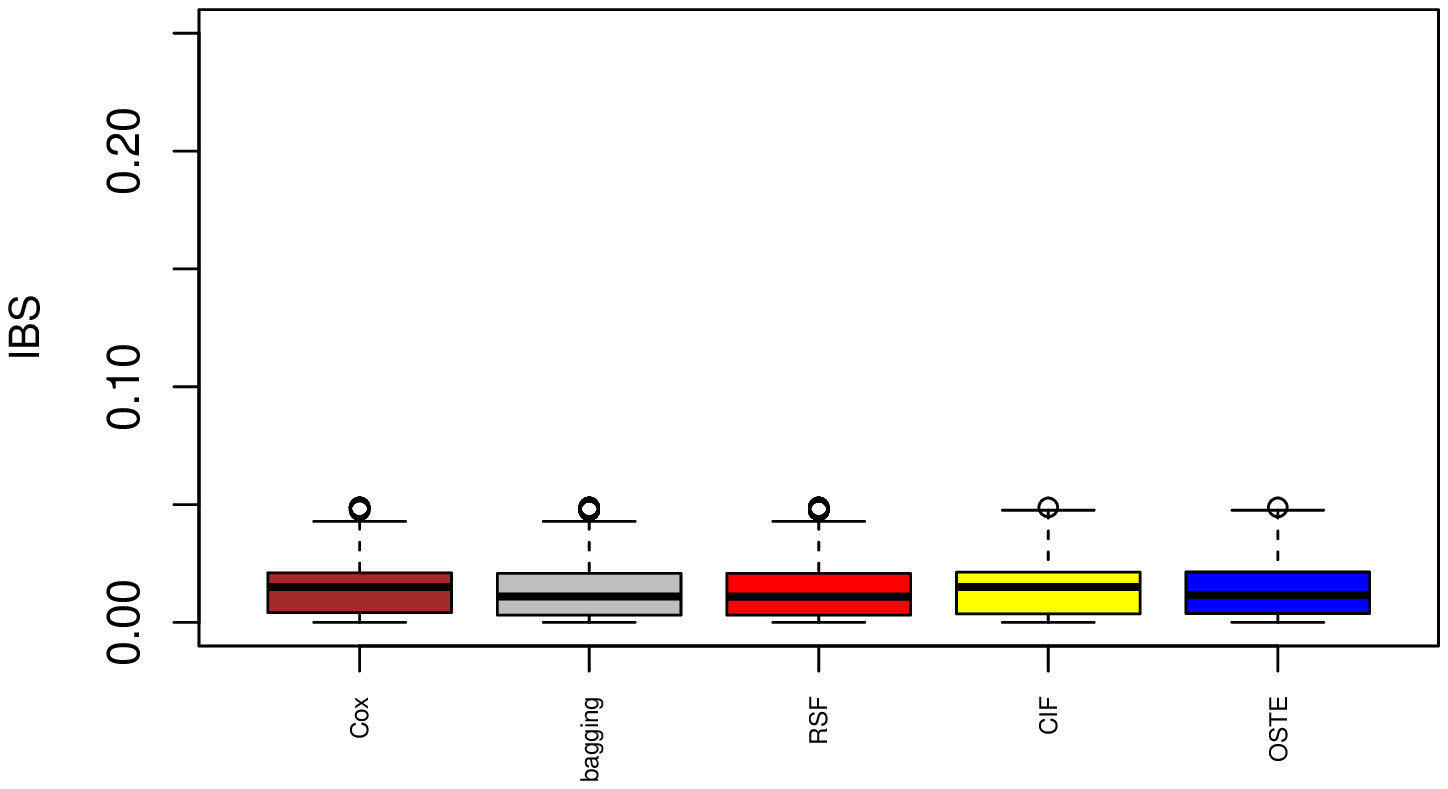} \\
	\text{ \ \ \ \ }(GBSG2)&\text{ \ \ \ \ }(burn)\\
	\includegraphics[width=6cm,height=5cm]{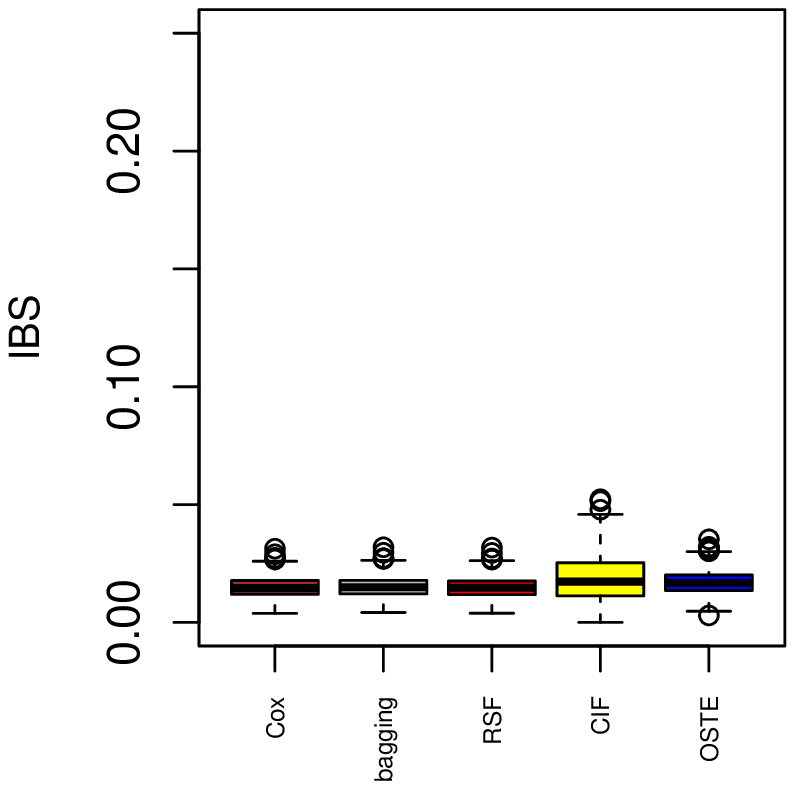} & \includegraphics[width=6cm,height=5cm]{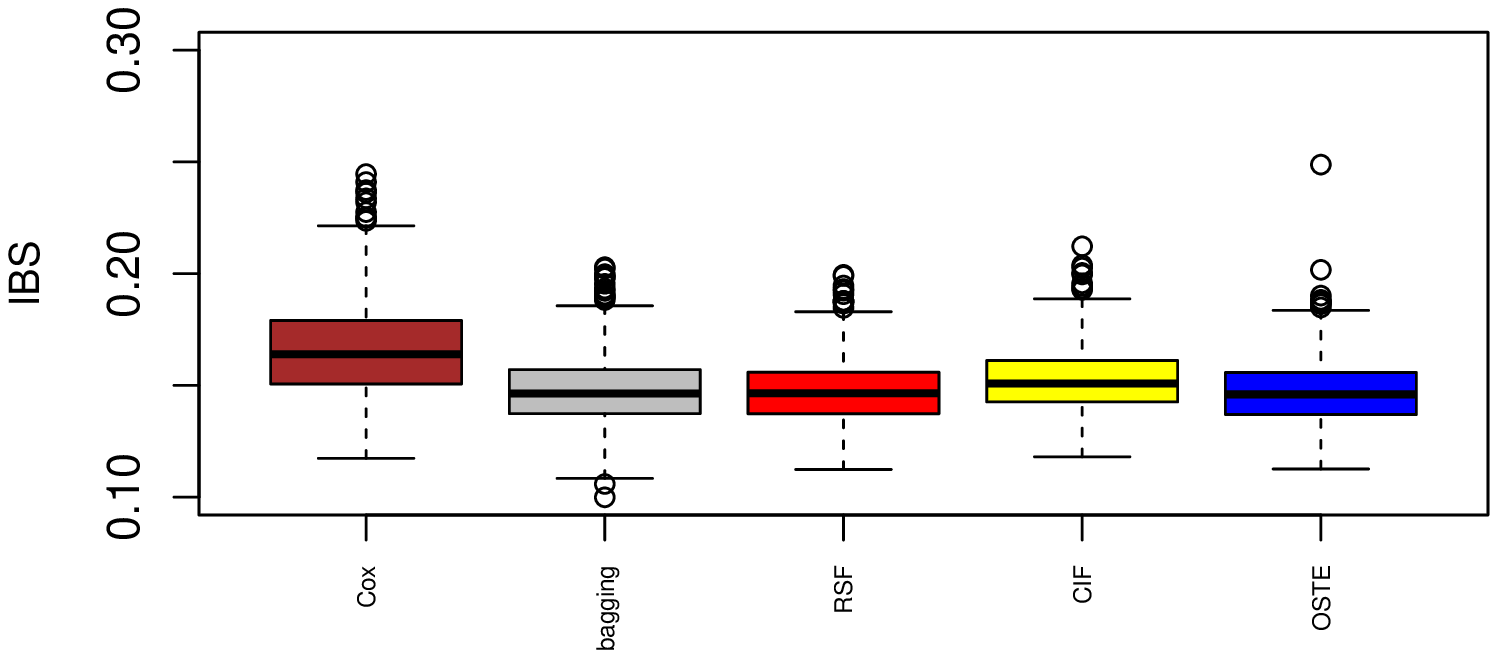} \\
	\end{array}$
	\label{set1}
	\caption{The boxplots showing IBS on the datasets veteran, kidtran, bfeed, twins, GBSG2 and burn. Cox, Bagging, RSF, CIF and OSTE are shown by brown, gray, red, yellow and blue colors, respectively. OSTE shows better performance on kidtran and bfeed datasets while for others datasets the results are similar to the alternative methods.}
\end{figure}

\begin{figure}[H]
	%\begin{figure}[h!]
	\centering
	$\begin{array}{cc}
	\text{ \ \ \ \ }(retinophty)&\text{ \ \ \ \ }(cgd)\\
	\includegraphics[width=5.9cm,height=5cm]{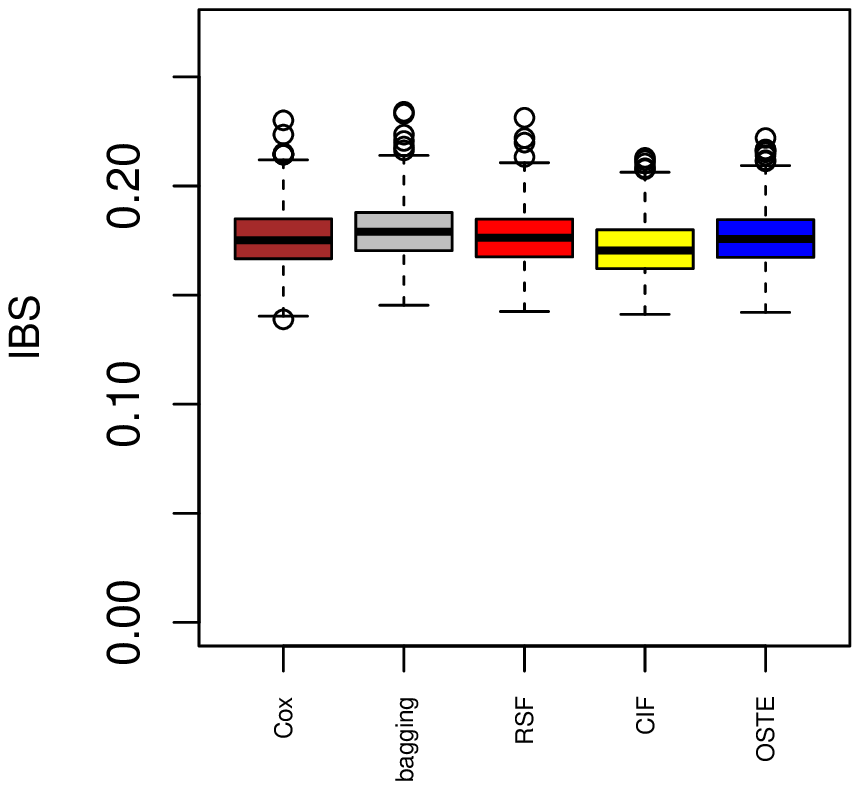} & \includegraphics[width=5.9cm,height=5cm]{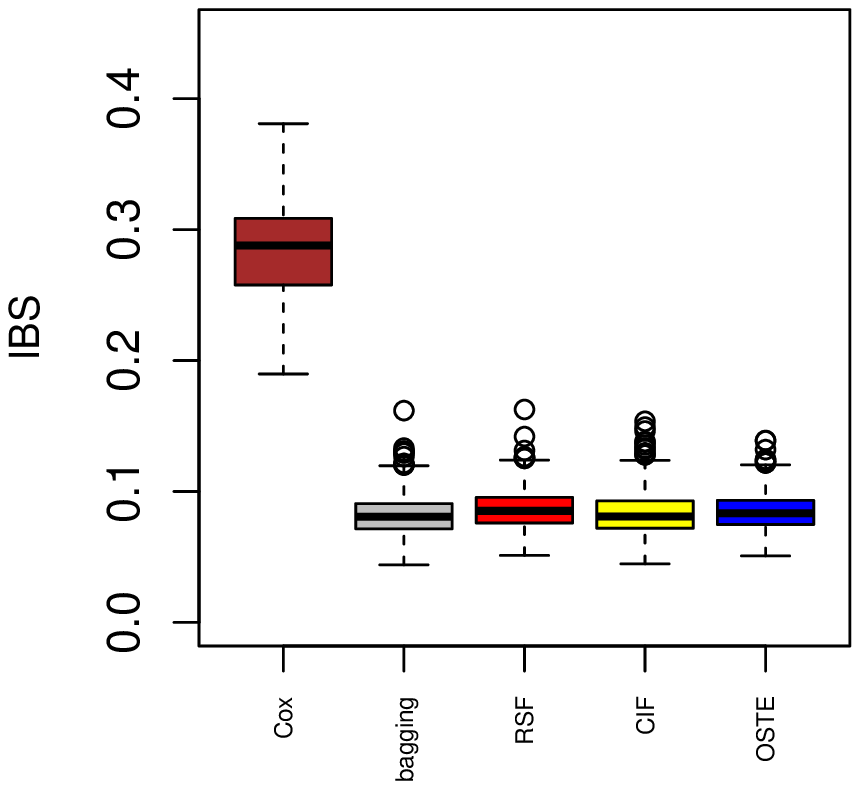} \\
	\text{ \ \ \ \ }(cost)&\text{ \ \ \ \ }(myeliod)\\
	\includegraphics[width=5.9cm,height=5cm]{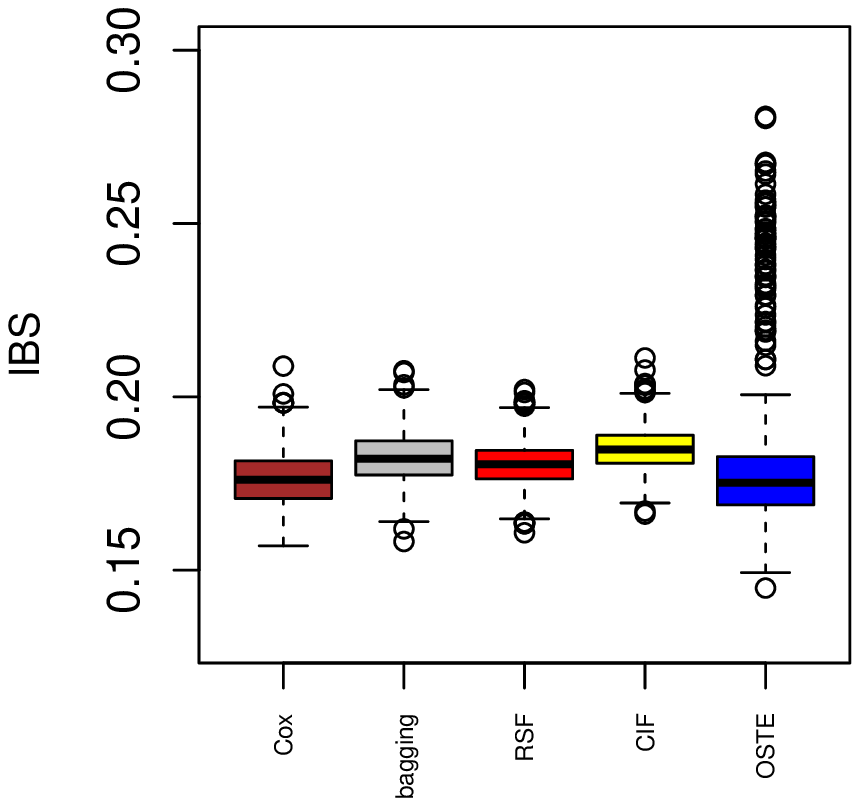} & \includegraphics[width=5.9cm,height=5cm]{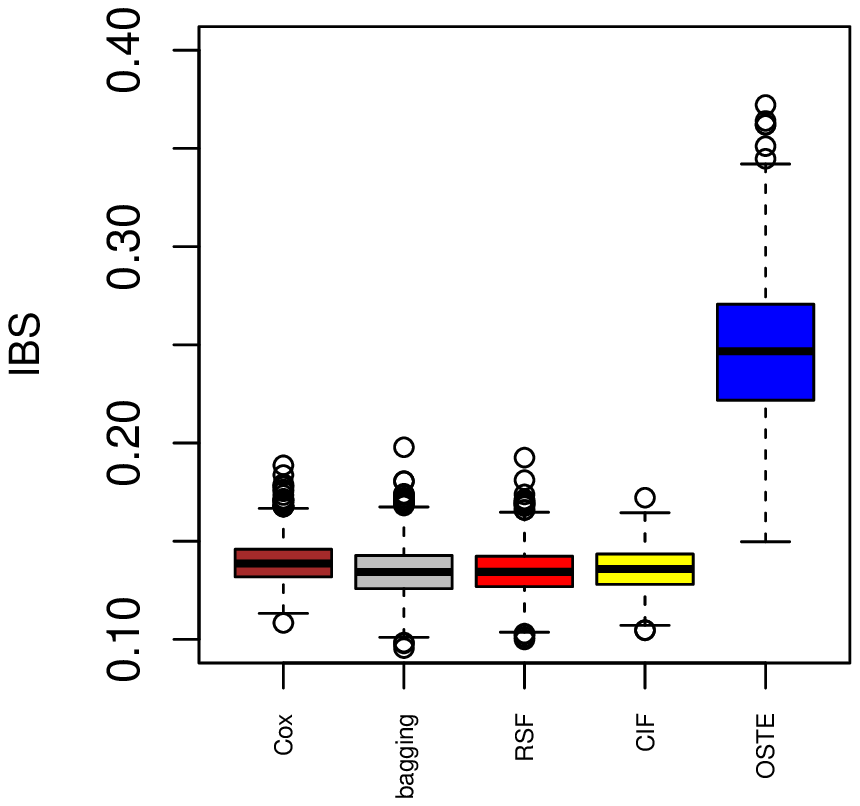} \\
	\text{ \ \ \ \ }(channing)&\text{ \ \ \ \ }(NKI)\\
	\includegraphics[width=5.9cm,height=5cm]{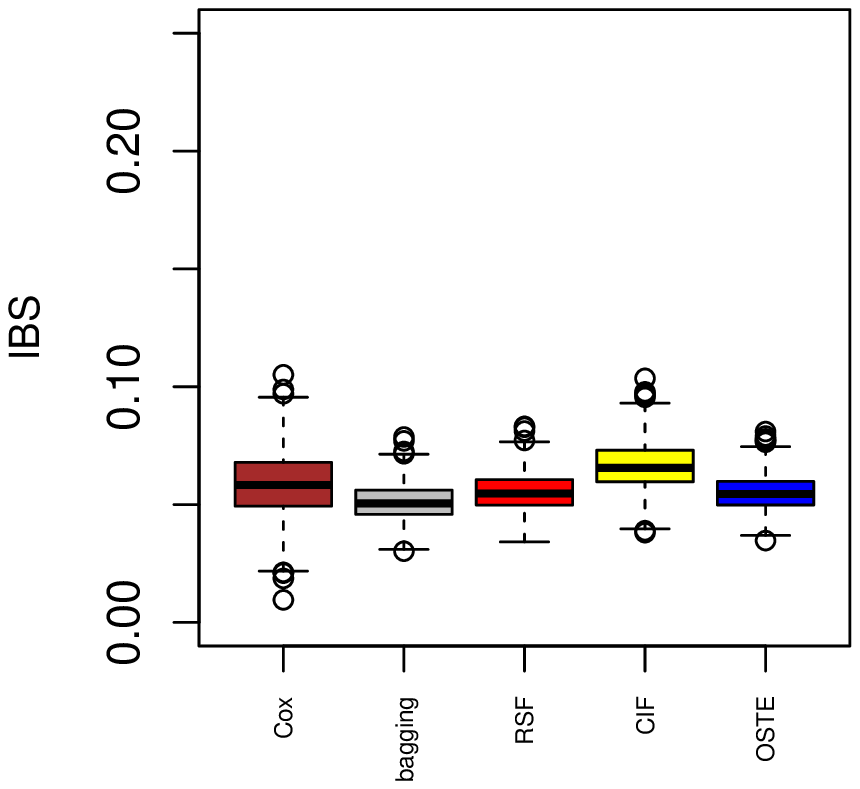} & \includegraphics[width=5.9cm,height=5cm]{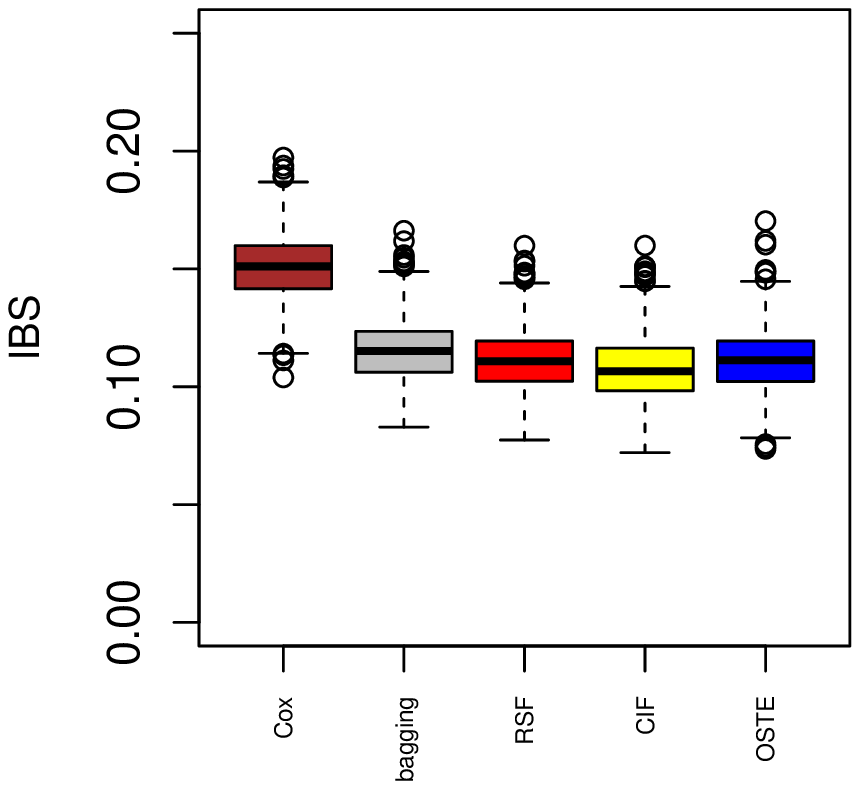} \\
	\end{array}$
	\label{set2}
	\caption{The boxplots showing IBS on the datasets retinophty, cgd, cost, myeliod, channing and NKI. Cox, Bagging, RSF, CIF and OSTE are shown by brown, gray, red, yellow and blue colors, respectively. OSTE shows similar performance on all the datasets except myeliod. }
\end{figure}
The boxplots given in Figure 3 show the IBS on the datasets retinophty, cgd, cost, myeliod, channing and NKI. Cox, Bagging, RSF, CIF and OSTE are shown by brown, gray, red, yellow and blue colors, respectively. The results of OSTE are almost same on all  datasets except myeliod. On cgd and NKI datasets the performance of Cox is poor while on other datasets the results of the methods are almost similar to the rest of the methods.
\begin{figure}[H]
	%\begin{figure}[h!]
	\centering
	$\begin{array}{cc}
	\text{ \ \ \ \ }(BMT)&\text{ \ \ \ \ }(colon)\\
	\includegraphics[width=6cm,height=5cm]{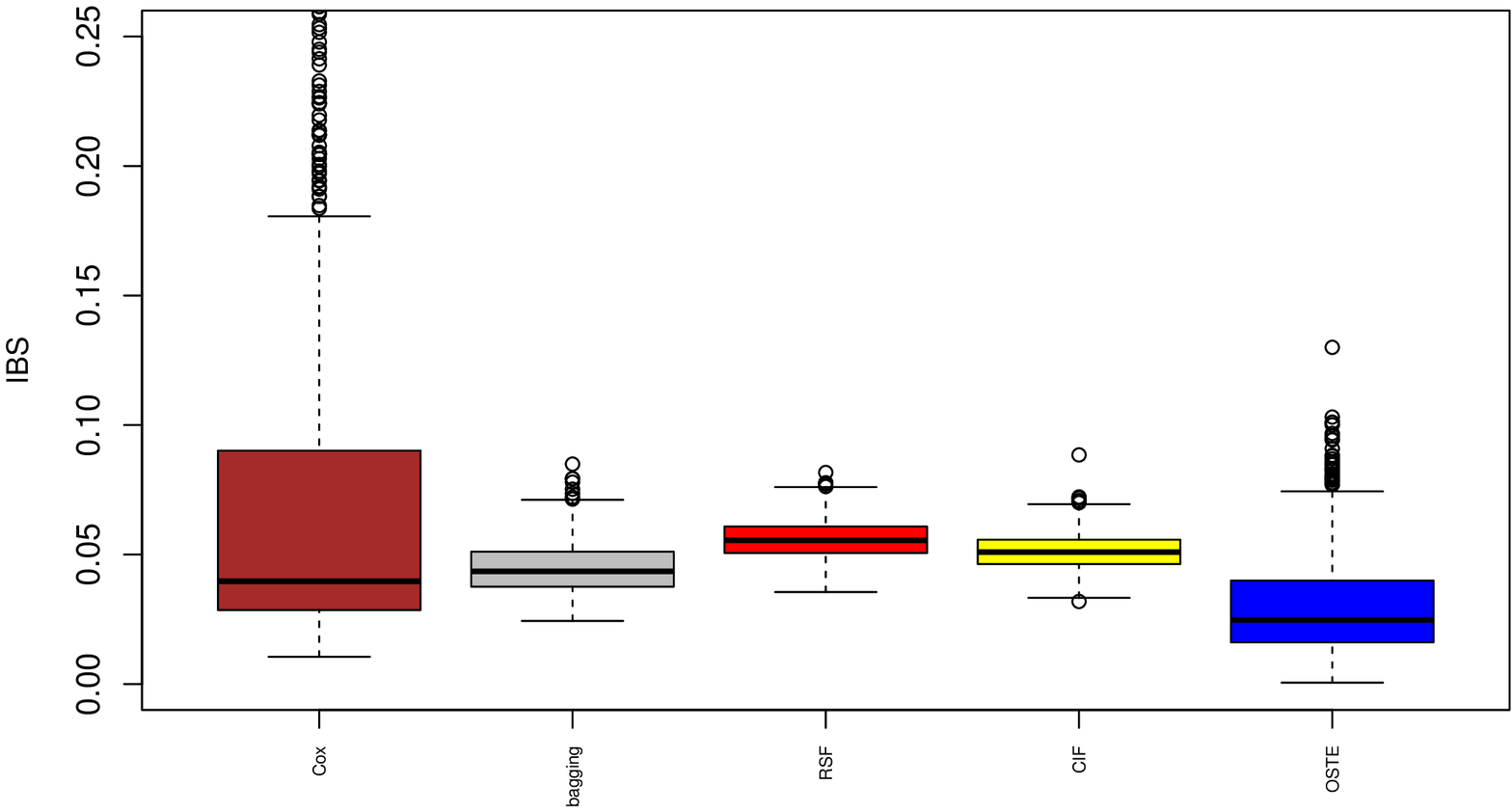} & \includegraphics[width=6cm,height=5cm]{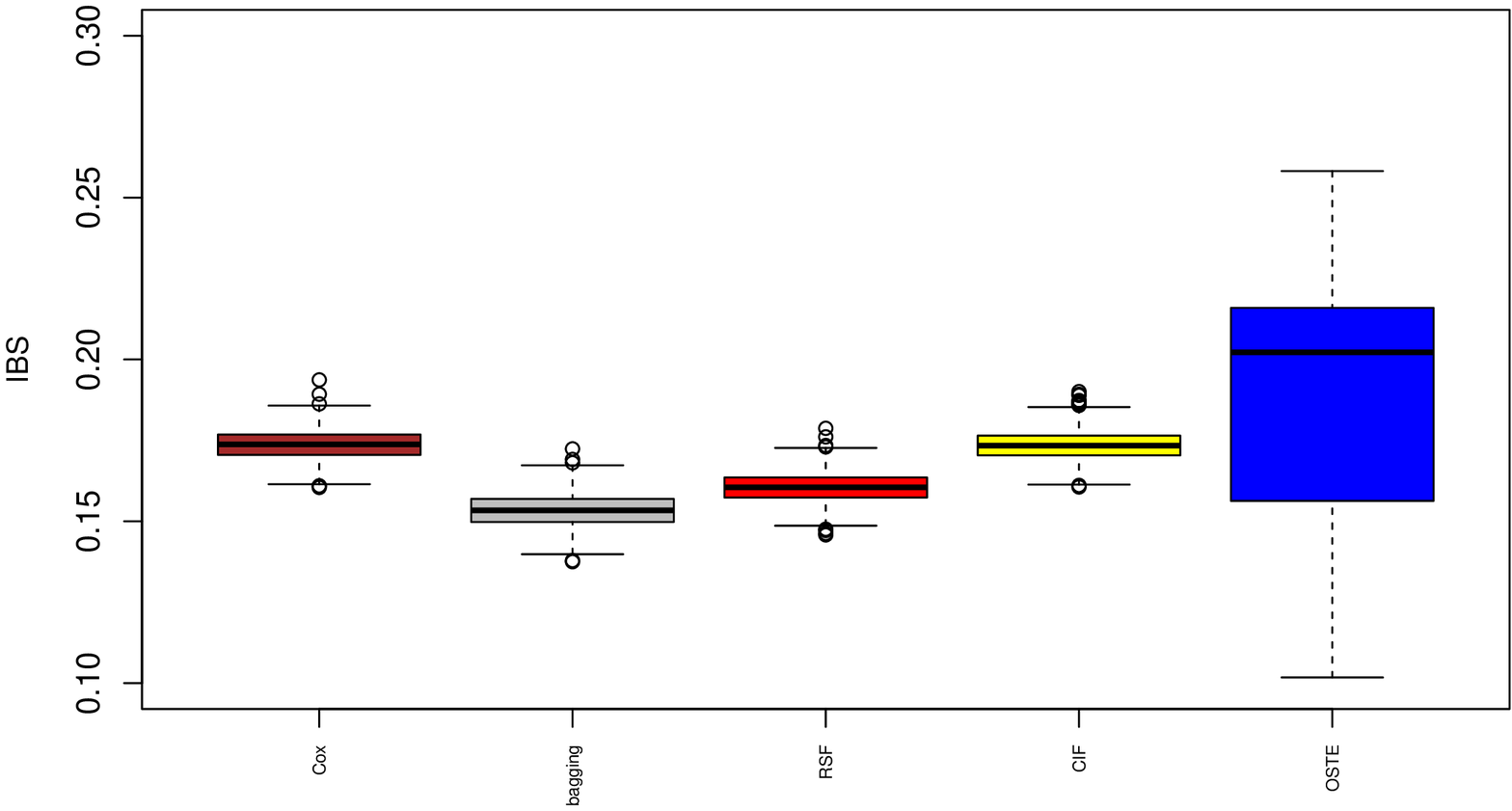} \\
	\text{ \ \ \ \ }(Hodg)&\text{ \ \ \ \ }(kidney)\\
	\includegraphics[width=6cm,height=5cm]{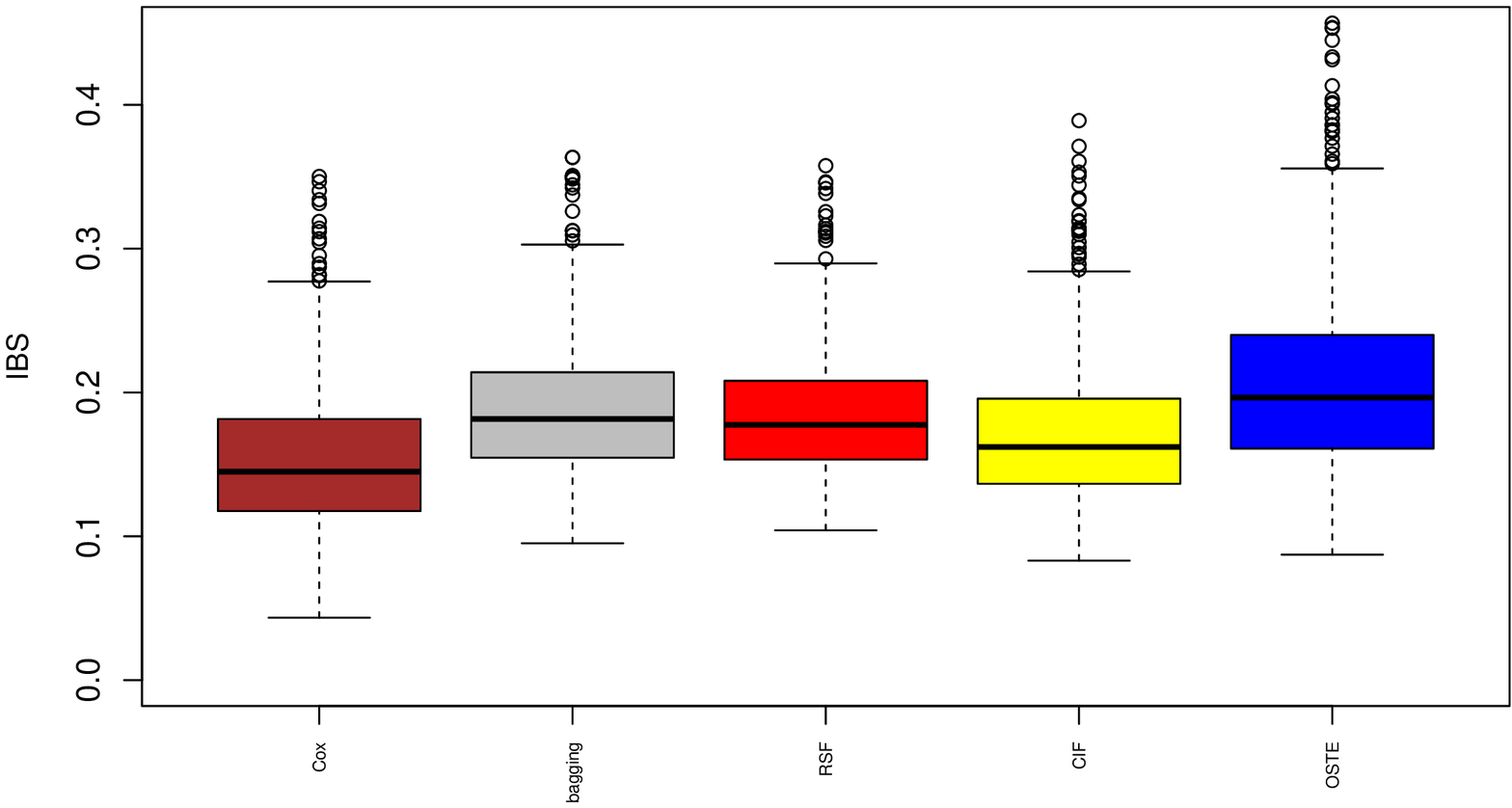} & \includegraphics[width=6cm,height=5cm]{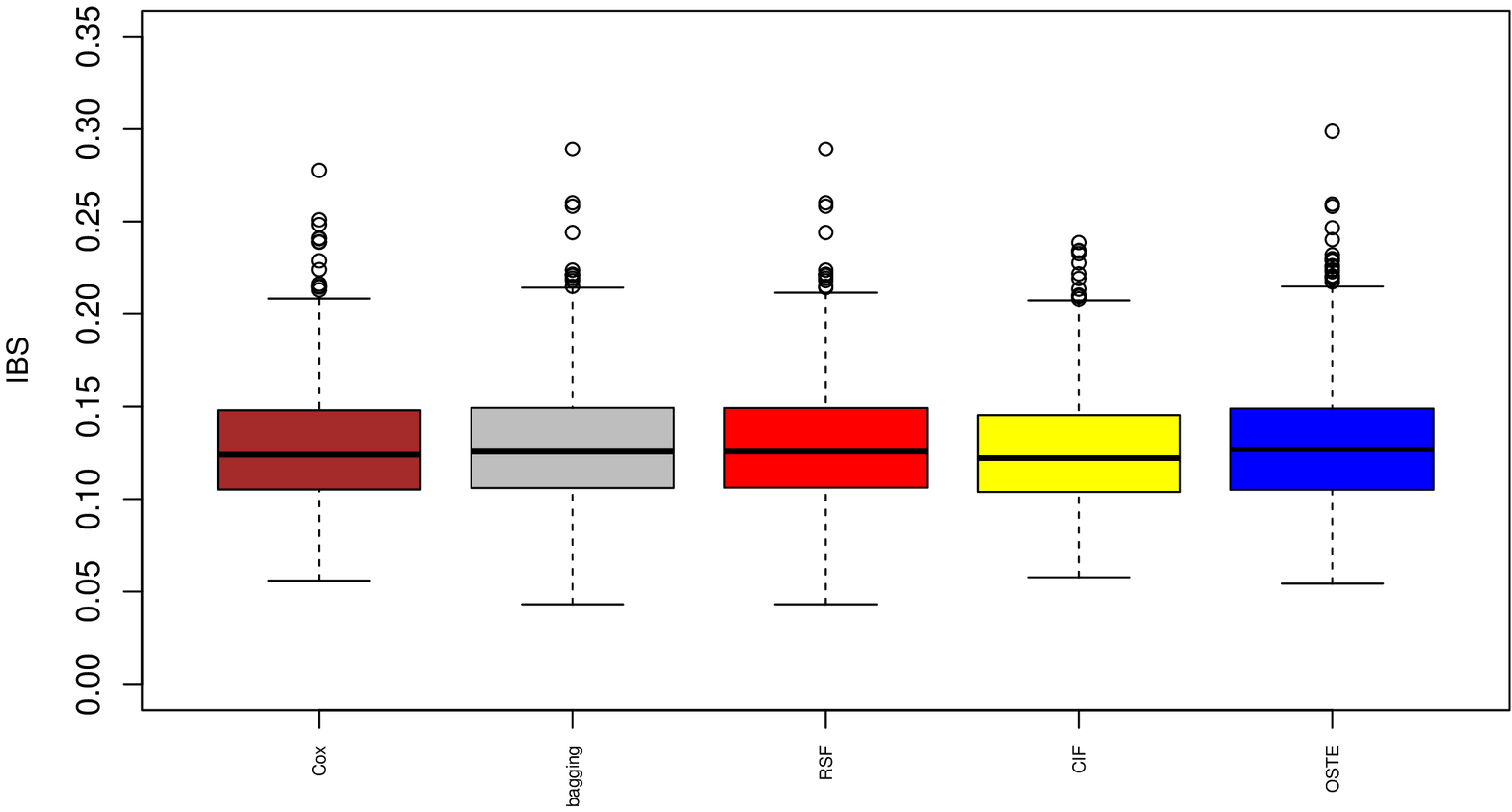} \\
	\text{ \ \ \ \ }(Pbc)\\
	\includegraphics[width=6cm,height=5cm]{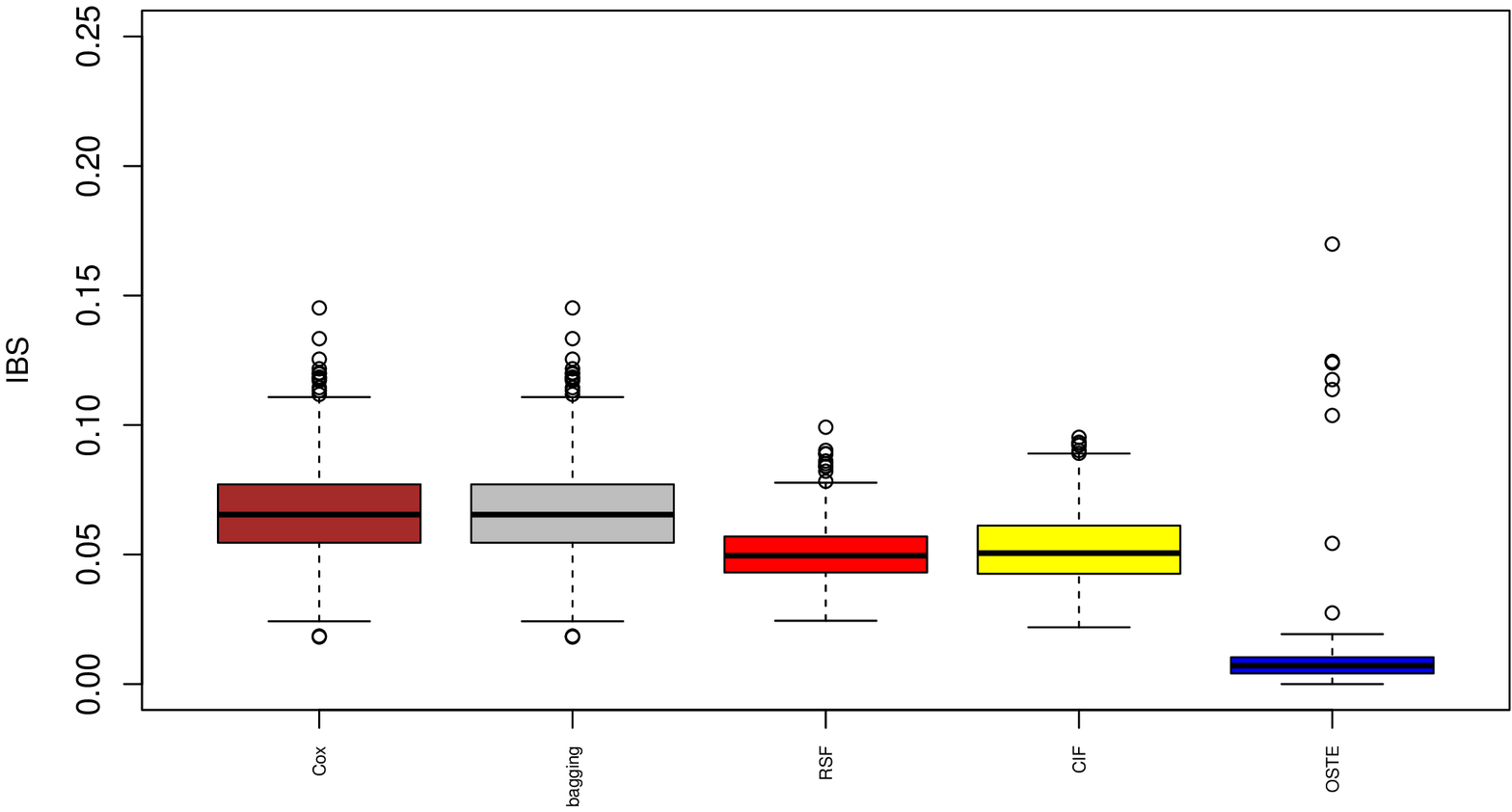}\\
	\end{array}$
	\caption{The boxplots showing IBS on the datasets BMT, colon, Hodg, kidney and Pbc. Cox, Bagging, RSF, CIF and OSTE are shown by  brown, gray, red, yellow and blue colors, respectively. For the kidney dataset OSTE shows similar performance while on Pbc and BMT datasets OSTE shows better results.}
	\label{set3}
\end{figure}
The boxplots given in Figure 4 showing IBS on the datasets BMT, colon, Hodg, kidney and Pbc. Cox, Bagging, RSF, CIF and OSTE are shown by  brown, gray, red, yellow and blue colours, respectively. For kidney dataset OSTE shows similar performance with the rest of methods while for Pbc and BMT datasets the results of OSTE are superior.

On some random splits of the data into training and testing parts OSTE might give comparatively larger error estimates as can be seen, for example in Figure 4.4 for Pbc data. This may happen when patterns in the selected trees are not in-line with those in the test data as OSTE selects trees with specific patterns.
%Due to tree selection with specific patterns, OSTE might give comparatively larger error estimates on some of the random splits of the data into training and testing parts. This may happen when patterns in the selected trees are not in-line with those in the test data. This can be seen in Figure 4.4 for Pbc data, for example.
Furthermore, a comparison of OSTE and RSF methods is given in terms of feature importance as OSTE improves RSF by removing trees from the original random survival forest with adverse effects on its overall efficiency. The permutation method 
%\cite{mielke2007permutation}
 is used fo this purpose. For survival tree, a given variable in the out-of-bag data is randomly permuted to estimate a variables permutation importance. After permutation, this OOB data is dropped down the tree and the OOB estimate of prediction error is calculated. 
%As OSTE improves RSF by discarding trees from the original forest with adverse effects on its overall efficiency, a further comparison of the two methods is given in terms of feature importance. Feature importance for both the methods is estimated via the permutation method \cite{ nicodemus2010behaviour }.For both the methods, a variable’s permutation importance is estimated by randomly permuting the given variable in the out-of-bag (OOB) data for the tree, and the permuted OOB data is dropped down the tree. The OOB estimate of prediction error is then calculated. 
 The estimate of the variable importance is the difference between this estimate and the OOB error without permutation, averaged over all trees. The larger the permutation importance of a variable, the more predictive the variable.

The estimate of the variable importance is checked on 4 data sets, burn, bmt, GBSG2 and colon for both the methods, OSTE and RSF as shown in Figure 4.5. It can be seen from the figure that for burn and bmt data sets OSTE give larger importance values to predictive features compared to random survival forest which might be due to the removal of harmful trees (i.e. the tree might have the effects of non-informative features) from the initially grown forest. For colon and GBSG2 datasets fails to give more importance to predictive features which might be the reason for the out performance of OSTE.  

%Variable importance on 4 data sets, burn, bmt, GBSG2 and colon is estimated for both the methods as shown in Figure 4.5. OSTE discards harmful trees from the forest that might have the effects of non-informative features thus giving larger importance values to predictive features compared to random survival forest as shown for the burn and bmt data sets (top panel of Figure 4.5). OSTE fails to achieve this in the cases of colon and GBSB2 data sets (bottom panel of Figure 4.5) which might be a reason of OSTE outperformed by RSF in the cases of these data sets. 
\begin{figure}[H]
	%\begin{figure}[h!]
	\centering
	$\begin{array}{cc}
	\includegraphics[width=6cm,height=5cm]{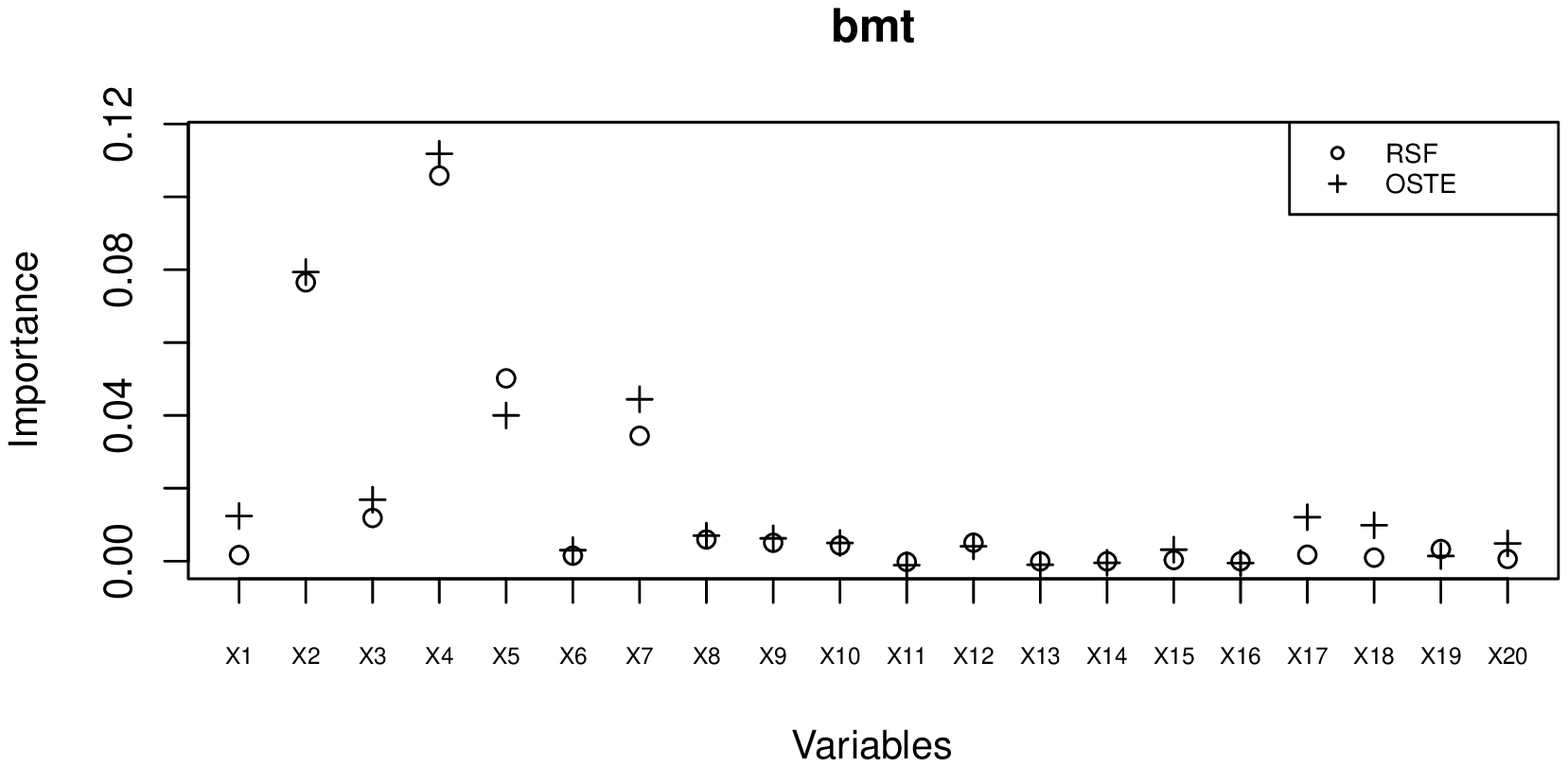} & \includegraphics[width=6cm,height=5cm]{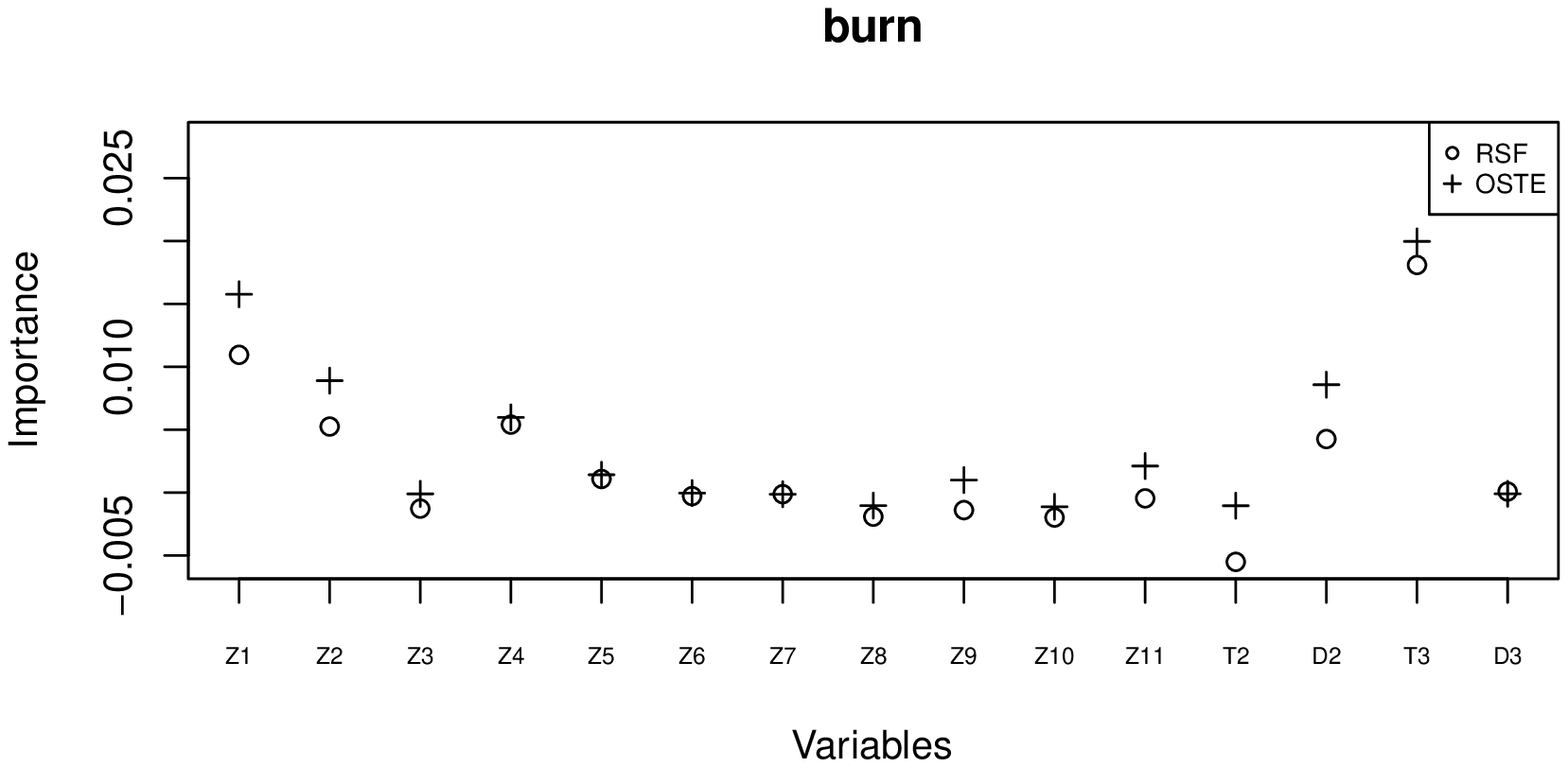} \\
	\includegraphics[width=6cm,height=5cm]{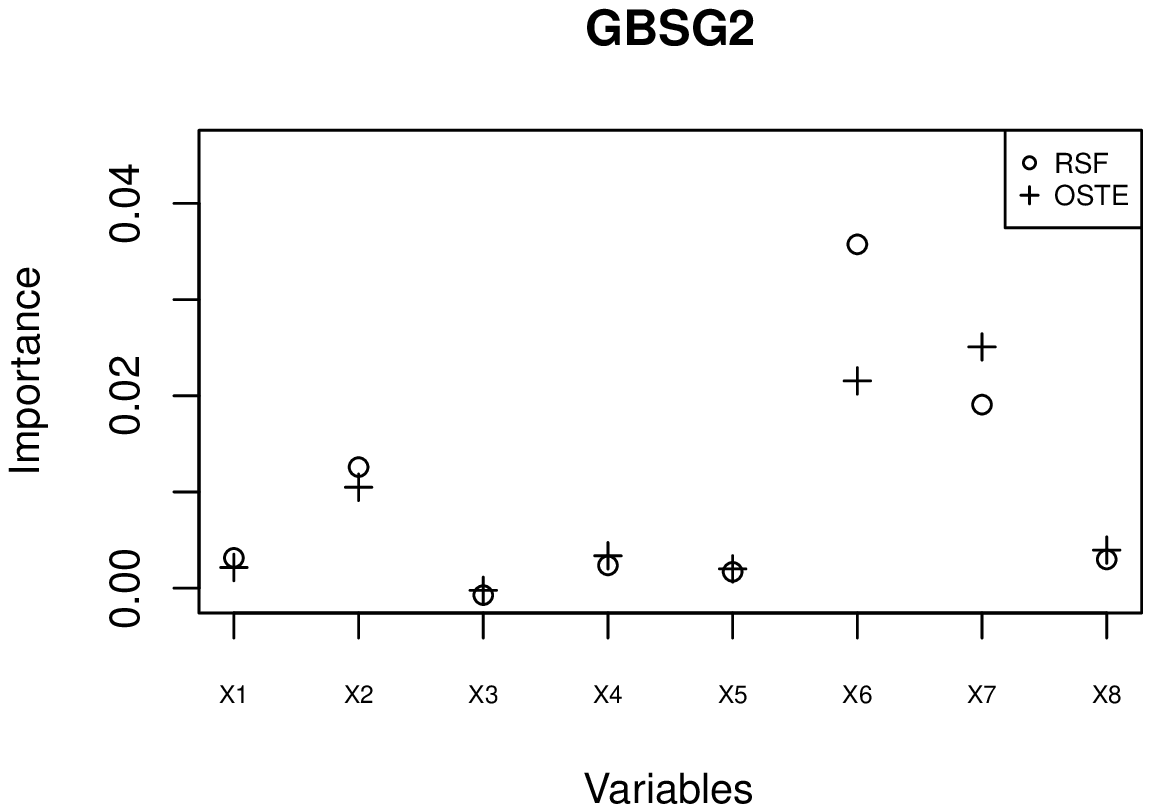} & \includegraphics[width=6cm,height=5cm]{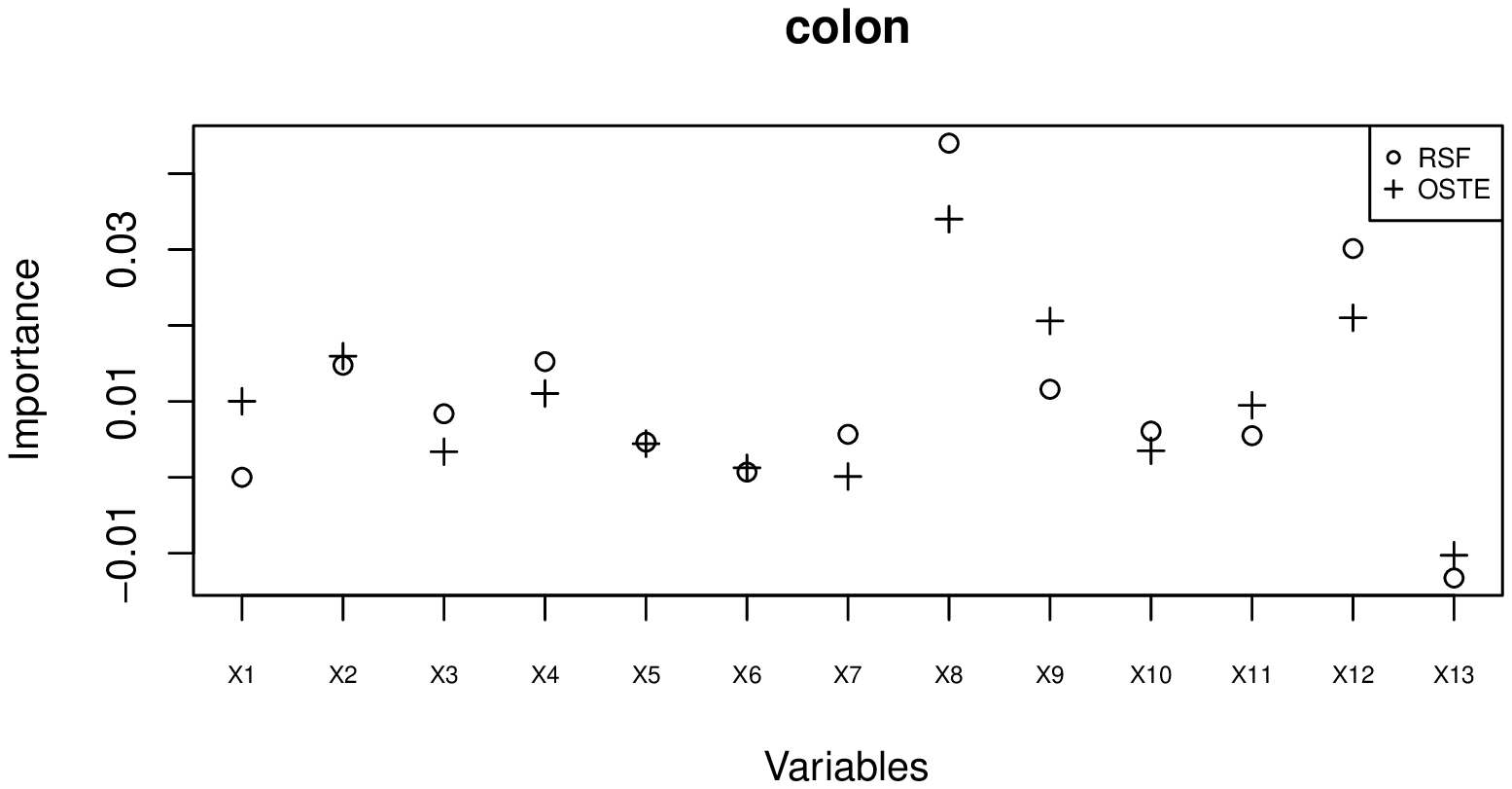} \\
	\end{array}$
	\caption{The plots showing feature importance for RSF and OSTE. The dots and + sign  shows RSF and OSTE respectively.} 
\end{figure}

\subsection{Hyper-parameters assessment}
\begin{figure}[H] 
	\centering 
	\includegraphics[width=14cm,height=10cm]{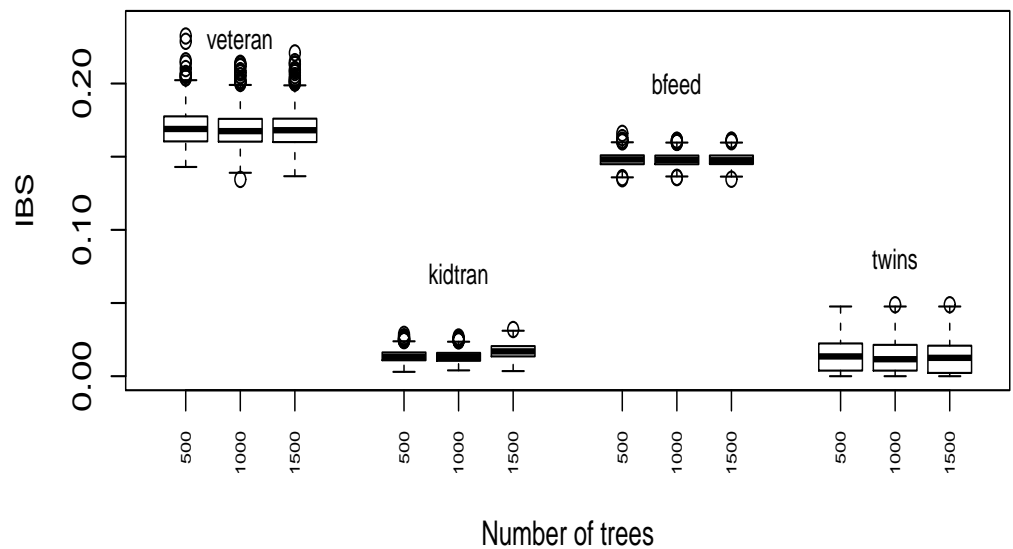}
	\caption{The boxplot showing a comparison of IBS on four datasets for different number of trees $B$ in the initial set.}
	\label{boxB}
\end{figure}

\begin{figure}[H]
	$\begin{array}{cc}
	\text{ \ \ \ \ }(veteran)&\text{ \ \ \ \ }(kidtran) \\
	\includegraphics[width=6cm,height=5cm]{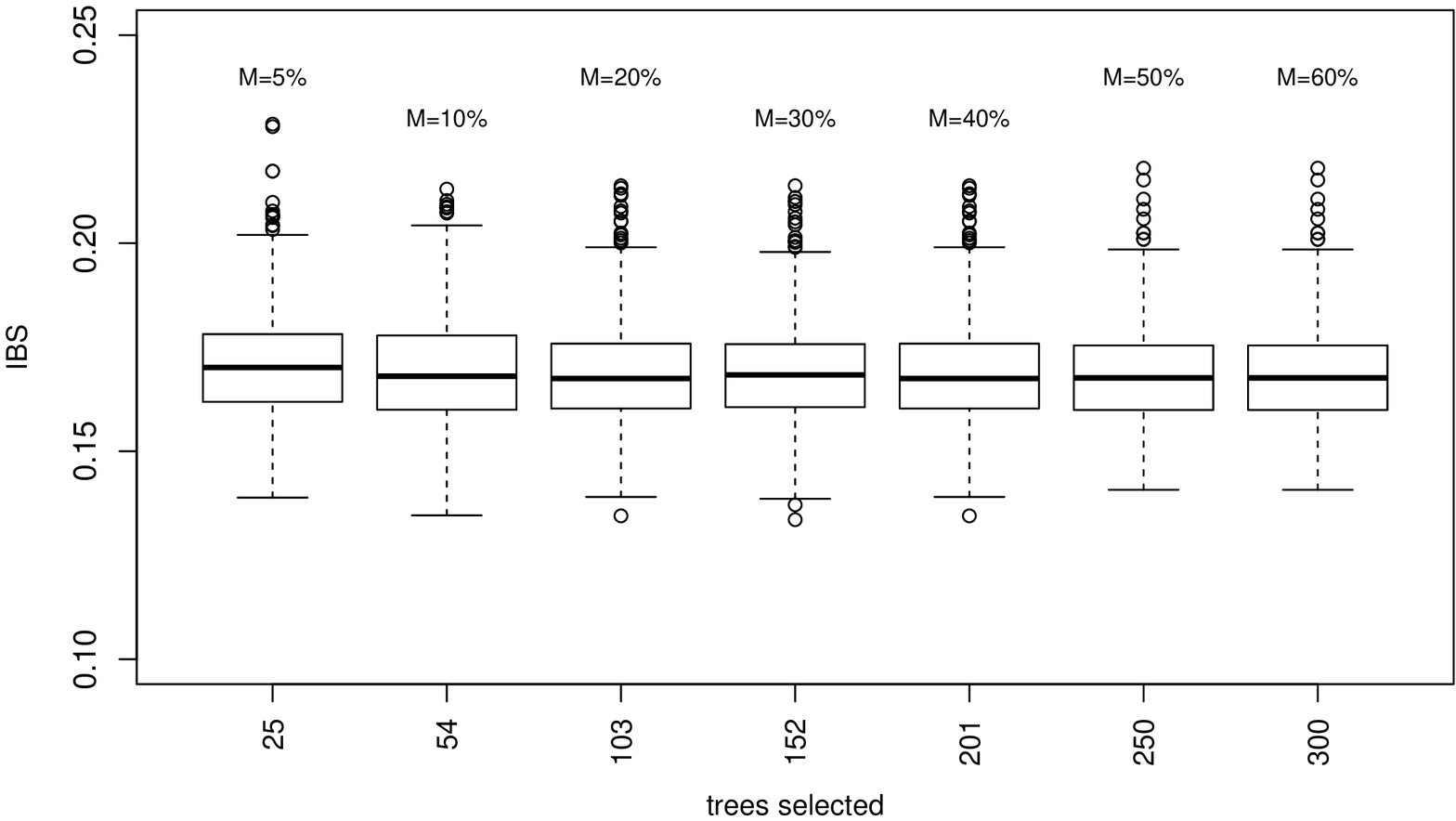} & \includegraphics[width=6cm,height=5cm]{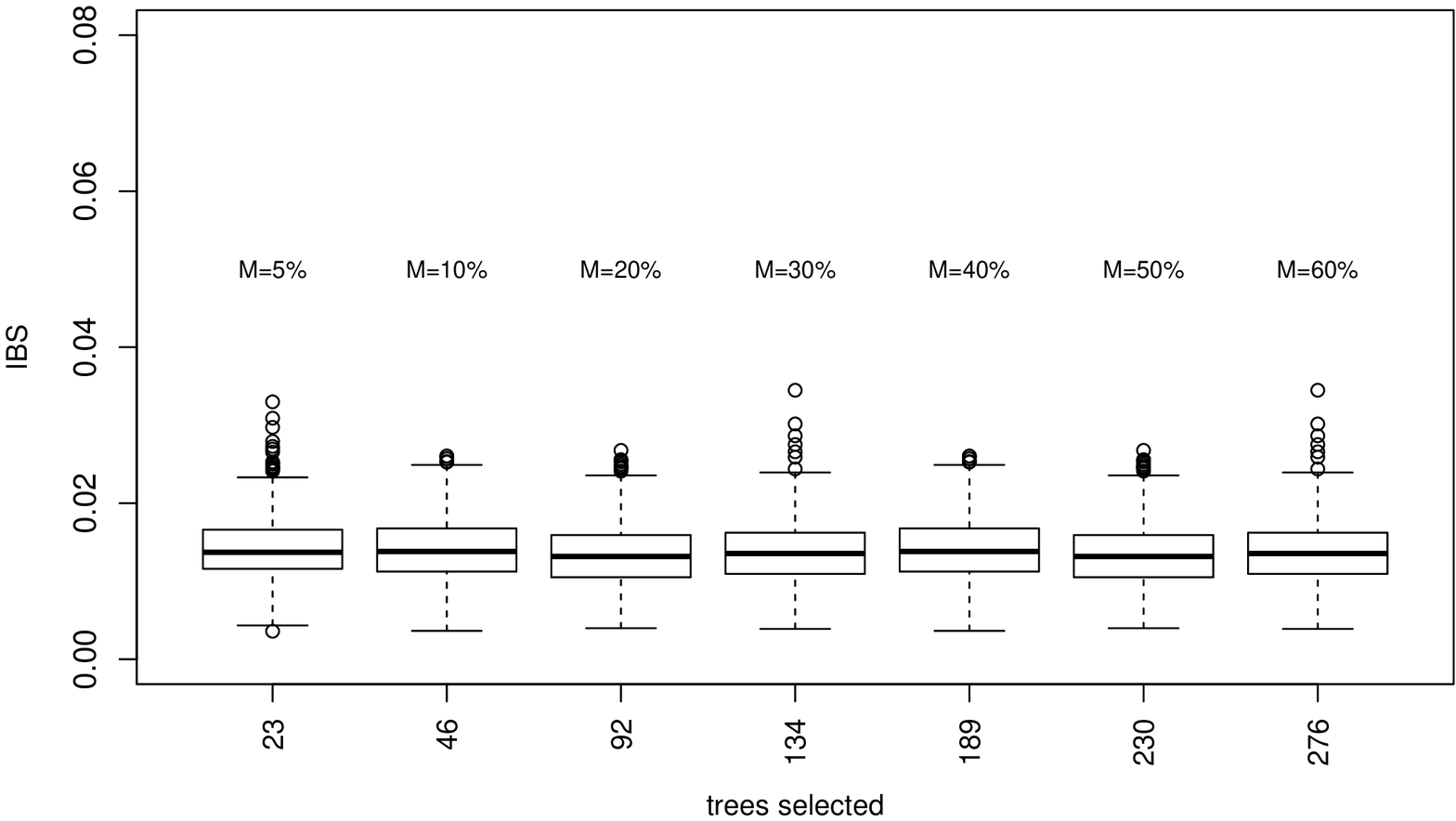}\\
	\text{ \ \ \ \ }(bfeed)&\text{ \ \ \ \ }(twins)\\
	\includegraphics[width=6cm,height=5cm]{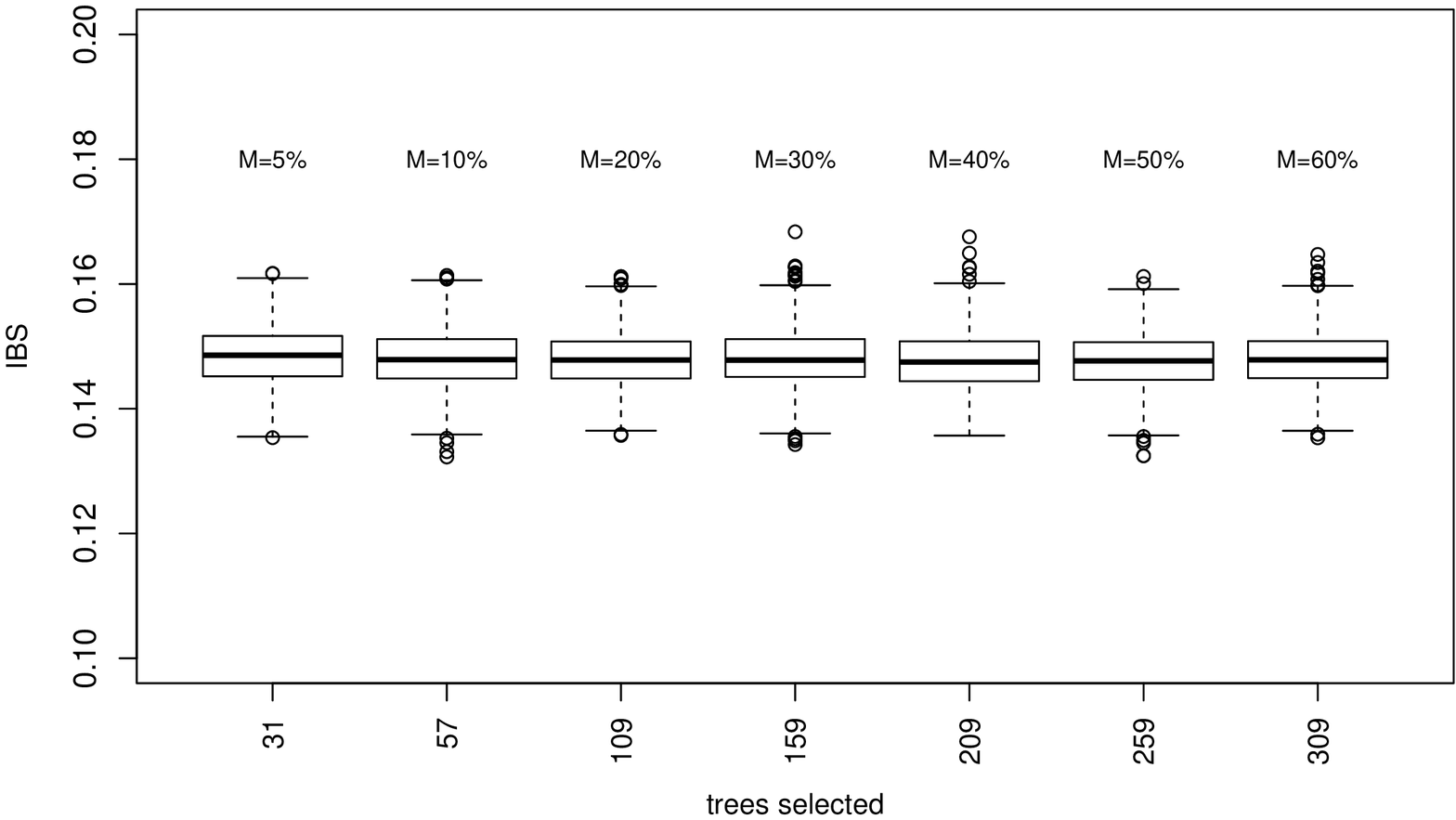} & \includegraphics[width=6cm,height=5cm]{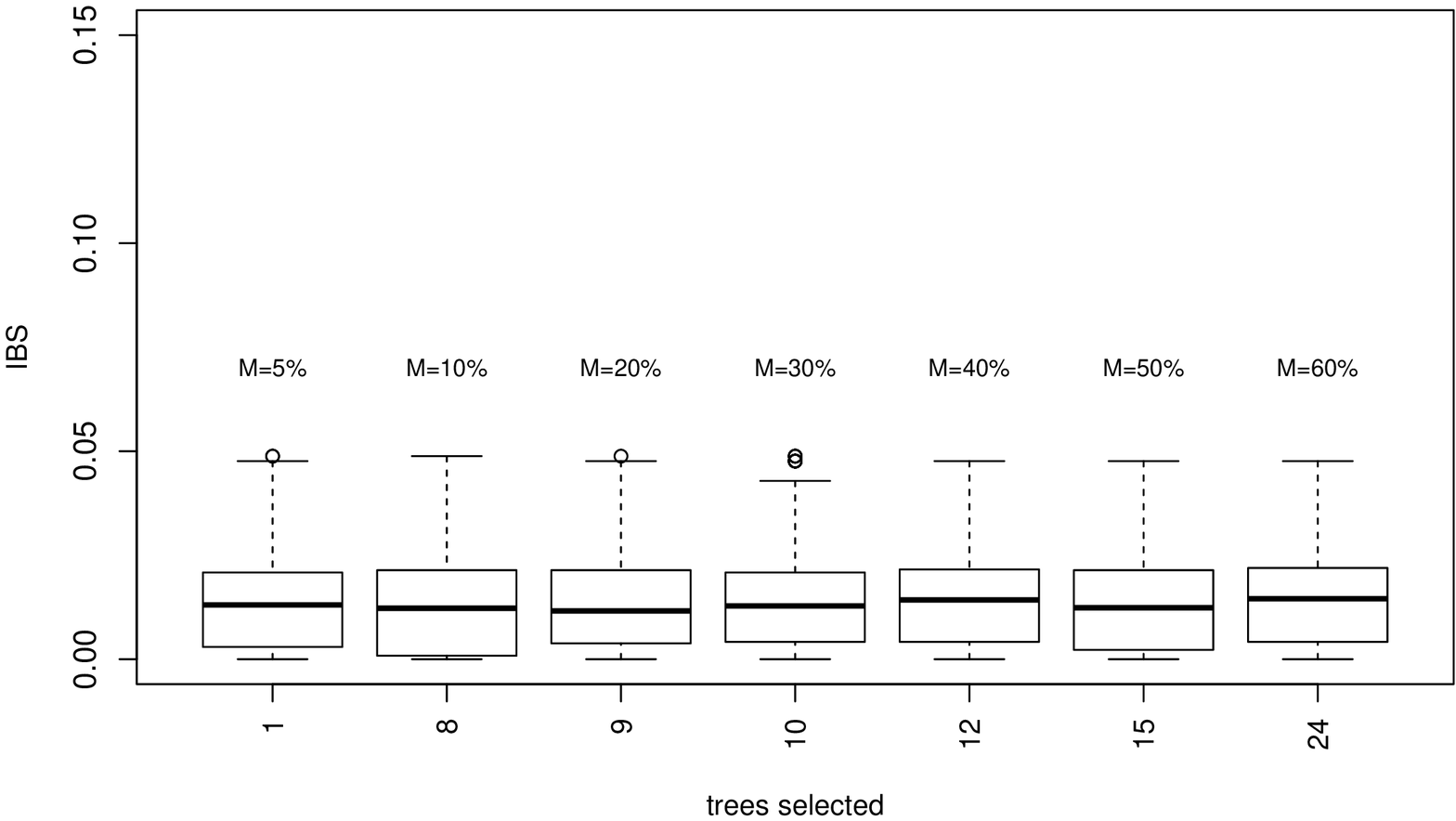} \\
	\end{array}$
	\caption{The boxplot showing a comparison of IBS on the datasets for different percentages of total number of trees ($M$) selected in the first phase. The trees selected by OSTE for the final ensemble are given on the x-axis.}
	\label{boxM}
\end{figure}

\begin{figure}[H]
	\centering 
	\includegraphics[width=14cm,height=10cm]{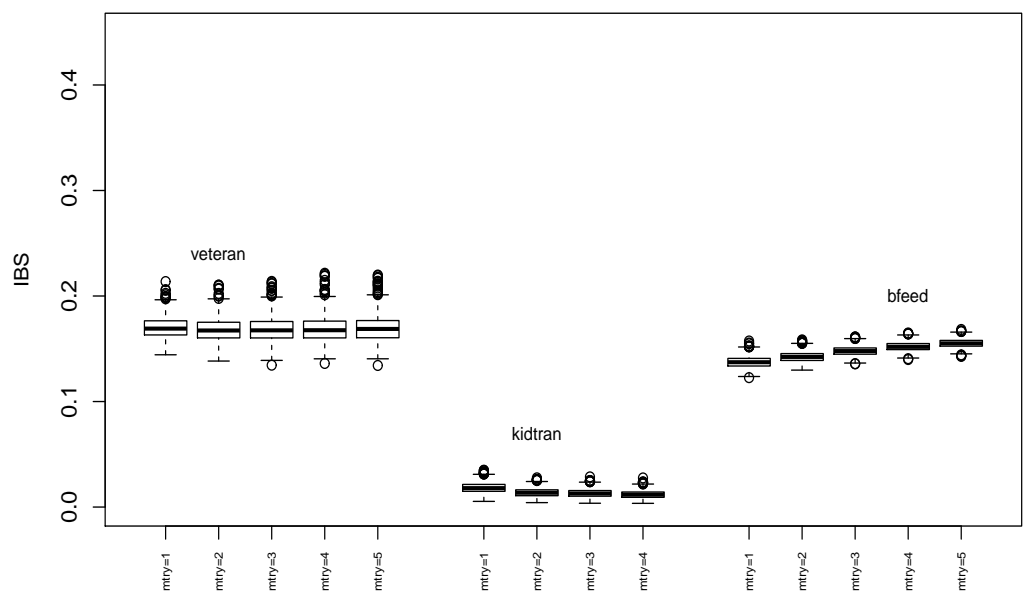}
	\caption{Boxplots showing a comparison of IBS on veteran, kidtran and bfeed datasets for different values of $p$.}
	\label{mtry}
\end{figure}

The effect of various number of  trees ($B$) grown initially in the ensemble, proportion of trees (M) chosen on individual accuracy and $p$ the number of features have been assessed on the results of the proposed method i.e. OSTE. The effect of $B$, is assessed on various values in initial set. The results are given in Figure \ref{boxB}. It can be seen that the Brier scores on the given datasets shows no/little effect with any increase in the number of trees from 1000, while for kidtran dataset, the error is increased by growing more than 1000 trees.
The proposed approach OSTE is also examined for various values of M i.e. $5\%, 10\%,\dots ,60\%$. The results are shown in Figure \ref{boxM}.
As can be seen from the results, OSTE shows same performance by only selecting 5\% of trees from the total grown trees initially. This selection of trees is based on individual accuracy against higher values of $M$ as shown in the figure. This has led to a resultant ensemble of sizes 25, 23, 31 and 24 for veteran, kidtran, bfeed and twins datasets respectively. This reveals that a significant reduction in the number of trees used for the final ensemble can be accomplished by using OSTE.
% \begin{figure}[h]
% 	%\begin{figure}[h!]
% 	\centering
% 	$\begin{array}{cc}
% 		\text{ \ \ \ \ }(veteran)&\text{ \ \ \ \ }(kidtran) \\
% 	\includegraphics[width=6cm,height=5cm]{./chapter4/veteran_M} & \includegraphics[width=6cm,height=5cm]{./chapter4/kidtran_M}\\
% 		\text{ \ \ \ \ }(bfeed)&\text{ \ \ \ \ }(twins)\\
% 	\includegraphics[width=6cm,height=5cm]{./chapter4/bfeed_M} & \includegraphics[width=6cm,height=5cm]{./chapter4/twins_M} \\
% 	\end{array}$
% 	\caption{The boxplot showing a comparison of IBS on the datasets for different percentages of total number of trees ($M$) selected in the first phase. The trees selected by OSTE for the final ensemble are given on the x-axis.}
% 	\label{boxM}
%\end{figure}
The effect of the number of features that we chose randomly for splitting the nodes of the trees on IBS are shown in Figure \ref{mtry}. As seen in the figure that for changing value of $p$ the results shows variations. The results suggest the tuning of this parameter for the corresponding data set.
% \begin{figure}[H]\label{mtry}
% 	\centering 
% 	\includegraphics[width=14cm,height=10cm]{./chapter4/different-mtry}
% 	\caption{Boxplots showing a comparison of IBS on veteran, kidtran and bfeed datasets for different values of $p$.}
% \end{figure}
\subsection{Size comparison}
In terms of the number of survival trees used a comparative analysis of ensemble sizes has also been done. The number of used survival trees in the resultant ensemble by the methods are given in Table \ref{size}. The table shows that a comparable performance could be achieved by selecting a total of 103, 92, 109, 1, 87, 35, 99, 104, 95, 102, 105, 203, 109, 97, 34, 46, 39 and 51 trees for veteran-Pbc datasets, receptively by selecting only $M=20\%$ as compared to the selection of hundreds of survival trees in the corresponding final ensembles by other state-of-the-art methods. This might be very helpful in reducing computational cost of the ensemble in terms of storage resources.
%The table shows that by selecting $M=20\%$, a comparable performance could be achieved by a total of 103, 92, 109, 1, 87, 35, 99, 104, 95, 102, 105, 203, 109, 97, 34, 46, 39 and 51 trees for veteran-Pbc datasets, receptively as compared to the other methods using hundreds of survival trees in the corresponding final ensembles. This might be very helpful in reducing computational cost of the ensemble in terms of storage resources. 

%\begin{table}[htbp]
%	\centering
%	\caption{Table showing sizes of ensemble for the datasets. Size of OSTE is shown for various values of $M$, taken in percentage.}
%	\begin{tabular}{lcccccccccc}
%		\toprule
%		\multicolumn{1}{c}{\multirow{2}[2]{*}{Dataset}} & \multicolumn{1}{c}{\multirow{2}[2]{*}{Bagging}} & \multicolumn{1}{c}{\multirow{2}[2]{*}{RSF}} & \multicolumn{1}{c}{\multirow{2}[2]{*}{CIF}} & \multicolumn{6}{c}{OSTE} \\
%		&       &       &       & \multicolumn{1}{l}{M=5} & \multicolumn{1}{l}{M=10} & \multicolumn{1}{l}{M=20} & \multicolumn{1}{l}{M=30} & \multicolumn{1}{l}{M=40} & \multicolumn{1}{l}{M=50}  & \multicolumn{1}{l}{M=60}\\
%		\midrule
%		veteran & 1000  & 1500  & 1000  & 25    & 54    & 103   & 152   & 201   & 250 & 300\\
%		kidtran & 1500  & 1000  & 1000  & 23    & 46    & 92    & 134   & 189   & 230 & 376\\
%		bfeed & 500   & 1000  & 500   & 31    & 57    & 109   & 159   & 209   & 259 & 309 \\
%		twins & 1000  & 1000  & 1500  & 4     & 8     & 9     & 10    & 12    & 15 & 24 \\
%		\bottomrule
%	\end{tabular}
%	\label{size}
%\end{table}

\begin{table}[htbp]
	\centering
	\caption{Table showing sizes of ensemble for the datasets. Size of OSTE is shown for $M = 20\%$.}
	\begin{tabular}{lccccccccc}
		\toprule
		Dataset & {Bagging} & {RSF} & {CIF}  & {OSTE} &       &       &  \\
		&       &       &       &&&&&&\\
		\midrule
		veteran & 1000  & 1500  & 1000  & 103   \\
		kidtran & 1500  & 1000  & 1000   & 92  \\
		bfeed & 500   & 1000  & 500  & 109   \\
		twins & 1000  & 1000  & 1500   & 1  \\
		VA   &  1000 &  1000  & 1500 & 87\\
		BMT  & 1000  & 1000  & 1000  & 35\\
		retinophty & 1000 &1000 &1000 & 99\\
		cgd & 1000 & 1000 & 1500 & 104\\
		channing & 1000 & 1500 &1000 & 95\\
		Burn &1500  &1000 &1500 &102\\
		GBSG2 &1500  &1500 &1500 & 105\\
		Cost  & 1500 &1000 &1000 & 203\\
		myeliod &1000 & 1000 &1500& 109\\
		NKI & 1000 & 1500& 1000& 97\\
		colon & 1500 &1500&1500&34\\
		Hodg & 1500 &1000 &1000& 46\\
		Kidney & 1000 & 1500& 1000 & 39\\
		Pbc & 1000 &1000 &1500 & 51\\
		\bottomrule
	\end{tabular}%
	\label{size}
\end{table}%
%\begin{table}[htbp]
%	\centering
%	\caption{Table showing sizes of ensemble for the datasets. Size of OSTE is shown for various values of $M$, taken in percentage.}
%	\begin{tabular}{lccccccccc}
%		\toprule
%		\multicolumn{1}{c}{\multirow{2}[2]{*}{Dataset}} & \multicolumn{1}{c}{\multirow{2}[2]{*}{Bagging}} & \multicolumn{1}{c}{\multirow{2}[2]{*}{RSF}} & \multicolumn{1}{c}{\multirow{2}[2]{*}{CIF}} & \multicolumn{6}{c}{OSTE} \\
%		&       &       &       & \multicolumn{1}{l}{M=5} & \multicolumn{1}{l}{M=10} & \multicolumn{1}{l}{M=20} & \multicolumn{1}{l}{M=30} & \multicolumn{1}{l}{M=40} & \multicolumn{1}{l}{M=50} \\
%		\midrule
%		veteran & 1000  & 1500  & 1000  & 25    & 54    & 103   & 152   & 201   & 250 \\
%		kidtran & 1500  & 1000  & 1000  & 23    & 46    & 92    & 134   & 189   & 230 \\
%		bfeed & 500   & 1000  & 500   & 31    & 57    & 109   & 159   & 209   & 259 \\
%		twins & 1000  & 1000  & 1500  & 4     & 8     & 9     & 10    & 12    & 15 \\
%		\bottomrule
%	\end{tabular}%
%	\label{size}%
%\end{table}%
\section{Conclusion}
The main aim of this paper is to lessen the number of survival trees in the final ensemble in addition to improving its performance. The idea of OSTE  ``optimal survival trees ensemble`` is proposed to achieve this goal.

To find trees that showed better performance based on the C-index, out-of-bag (OOB) observations from the bootstrap samples are used as the test subjects. Ensemble predictive accuracy is checked by assessing top ranked survival trees on independent training data. For the final ensemble those survival trees who performed well both individually and collectively have been selected. The results, in terms of the integrated Brier score (IBS), of 17 benchmark datasets after applying OSTE, are compared with Cox proportional hazard model, random survival forest, conditional inference forest and bagging survival trees. 

% OSTE is then applied on 17 benchmark datasets and the results, in terms of integrated Brier score (IBS), are compared with  some state-of-the-art methods i.e. Cox proportional hazard model, random survival forest, conditional inference forest and bagging survival trees. 

From the integrated Brier scores, after applying all the methods to the datasets discussed above, average IBS values are calculated. It has been observed from the final results in term of boxplots, that OSTE, the proposed method, is giving better or comparable results to the best of the other methods.
%Average IBS are calculated and Boxplots have been constructed from the integrated Brier scores after applying all the methods on the aforementioned datasets. It has been observed that OSTE, the proposed method, is giving better/comparable results to the best of the other methods.

It has also been observed that OSTE reduced the number of survival trees in the resultant ensemble and improved predictive performance. OSTE consisting of less than 20 survival trees is seen to give comparable results to those ensembles which select hundreds of survival trees.

Furthermore, OSTE  performance has also been checked for various hyper-parameters. In this regard, the effect of changing the selected features number i.e, $p$ at the nodes of the survival trees, number of trees in the initial ensemble and proportion $M$ of the top ranked trees have been assessed. $p$ is considered as a tuning parameter of the proposed method and should be fine tuned for a given dataset accordingly. $M$ needs not to be greater than 20\% as higher values of M show no improvements and only increase the size of the ensemble. A total of 1000 survival trees are suggested to be grown in an initial set, for best results. 
The proposed ensemble is implemented in an $R$ package ``OSTE``. 

For the purposes of internal validation, OSTE leaves some observations from the training data during bootstrapping, therefore, some information is lost in the learning process. Whereas the remaining methods use the whole training set. Thus the performance of OSTE may be negatively affected. These out-of-bag observations could be used for internal validation as a future direction for research into this problem and further improve OSTE. 

While growing a survival tree, a maximally selected rank statistic \cite{lausen1992maximally} could be used instead of the log-rank statistic for the selection of split points as the log-rank test favours many split points. 
%Another way to further improve the proposed method is to look for alternative statistics instead of log-rank test while growing survival tree in that this test favours splitting on variables with many possible split points. Maximally selected rank statistic \cite{lausen1992maximally} could serve as an off-the-shelve tool for split point selection while growing trees for OSTE.  

For the proposed method to work well in high dimensional settings, some state-of-the-art feature selection/dimensionality reduction techniques could be used with OSTE. 
\bibliographystyle{spmpsci}      % mathematics and physical sciences
\bibliography{ostebib}   % name your BibTeX data base
\end{document}